\documentclass[journal]{IEEEtran}
\usepackage{amsmath,amssymb,amsfonts}
\usepackage{float}
\usepackage{graphicx}
\usepackage{epstopdf} 
\usepackage{textcomp}
\usepackage{cite}

\usepackage{mathtools}
\usepackage[ruled,linesnumbered,noend]{algorithm2e}
\usepackage{caption}
\usepackage{tabularray}
\usepackage{multirow}
\usepackage{subcaption}
\usepackage{booktabs}
\usepackage{enumerate}
\usepackage{rotating}
\usepackage{bbding}
\usepackage{xspace}
\usepackage{balance}

\usepackage{tikz}

\usepackage{xcolor,colortbl}
\usepackage{tcolorbox}

\usepackage{bm}
\usepackage[nolist,nohyperlinks]{acronym}
\usepackage[noend]{algpseudocode}
\usepackage[inline]{enumitem}
\usepackage{url}
\usepackage{authblk}

\def\checkmark{\tikz\fill[scale=0.4](0,.35) -- (.25,0) -- (1,.7) -- (.25,.15) -- cycle;} 
\let\oldReturn\Return
\renewcommand{\Return}{\State\oldReturn}

\definecolor{green}{rgb}{0.1,0.1,0.1}

\graphicspath{ {./} }
\tcbuselibrary{theorems}
\newtcbtheorem
  []
  {takeaway}
  {Takeaway}
  {%
    colback=green!5,
    colframe=green!35!black,
    fonttitle=\bfseries,
  }
  {tay}

\usepackage[nolist,nohyperlinks]{acronym}

\newcommand{\fedavg}{\textit{FedAvg}\xspace}
\newcommand{\fedprox}{\textit{FedProx}\xspace}
\newcommand{\fedbn}{\textit{FedBN}\xspace}
\newcommand{\newadd}[1]{\textcolor{black}{#1}}

\begin{document}

\title{Benchmarking Federated Learning for Throughput Prediction in 5G Live Streaming Applications}

\author{Yuvraj Dutta, Soumyajit Chatterjee,~\IEEEmembership{Member,~IEEE}, Sandip Chakraborty,~\IEEEmembership{Senior~Member,~IEEE,} Basabdatta Palit,~\IEEEmembership{Member,~IEEE}
\thanks{Y. Dutta and B.Palit are with the Department
of Electronics and Communication Engineering, National Institute of Technology Rourkela, India - 769008. S. Chatterjee is with Nokia Bell Labs, UK and University of Cambridge, UK.  S. Chakraborty is with the Department of Computer Science and Engineering, Indian Institute of Technology Kharagpur, India 721302. Emails: duttayuv.nitrkl@gmail.com, soumyajit.chatterjee@nokia-bell-labs.com, palitb@nitrkl.ac.in, sandipc@cse.iitkgp.ac.in.}}



\maketitle

\begin{abstract}

Accurate and adaptive network throughput prediction is essential for latency-sensitive and bandwidth-intensive applications in 5G and emerging 6G networks. However, most existing methods rely on centralized training with uniformly collected data, limiting their applicability in heterogeneous mobile environments with non-IID data distributions.

This paper presents the first comprehensive benchmarking of federated learning (FL) strategies for throughput prediction in realistic 5G edge scenarios. We evaluate three aggregation algorithms - \textit{FedAvg}, \textit{FedProx}, and \textit{FedBN} - across four time-series architectures: LSTM, CNN, CNN+LSTM, and Transformer, using five diverse real-world datasets. We systematically analyze the effects of client heterogeneity, cohort size, and history window length on prediction performance.

Our results reveal key trade-offs among model complexities, convergence rates, and generalization. It is found that \textit{FedBN} consistently delivers robust performance under non-IID conditions. On the other hand, LSTM and Transformer models outperform CNN-based baselines by up to 80\% in $R^2$ scores. Moreover, although Transformers converge in half the rounds of LSTM, they require longer history windows to achieve a high $R^2$, indicating higher context dependence. LSTM is, therefore, found to achieve a favorable balance between accuracy, rounds, and temporal footprint.

To validate the end-to-end applicability of the framework, we have integrated our FL-based predictors into a live adaptive streaming pipeline. It is seen that \textit{FedBN}-based LSTM and Transformer models improve mean QoE scores by 11.7\% and 11.4\%, respectively, over \textit{FedAvg}, while also reducing the variance. These findings offer actionable insights for building scalable, privacy-preserving, and edge-aware throughput prediction systems in next-generation wireless networks.

\end{abstract}


\section{Introduction}\label{sec:intro}
\IEEEPARstart{T}{he} increasing demand for high-bandwidth, low-latency applications in next-generation wireless networks, such as 5G and the emerging 6G, has made accurate and robust network throughput prediction indispensable for sustaining performance under dynamic and resource-constrained network conditions. A wide range of latency-sensitive applications, such as ultra-HD and 360$^\circ$ video streaming, immersive augmented and virtual reality (AR/VR), and real-time digital twins, place stringent demands on the underlying network infrastructure. These applications are inherently interactive, with continuous data exchange between the edge servers and the client devices, often across heterogeneous and mobile environments. In such settings, the underlying network conditions may vary rapidly due to mobility, interference, congestion, or changes in user densities. Since these applications typically operate close to the limits of available network capacity, predicting the future throughput accurately becomes critical for maintaining responsiveness, minimizing disruptions, and ensuring a consistently high \ac(QoE)~\cite{VidQoEVenkatraman2007,VidQoEVenkataraman2009,VidQoEssaili2013,VidQoEYitong2013,VidQoEBarman2019,ABRstrYusuf2007,LSQoESun2019,LSQoEWei2021,ARHirway2020,ARWallace2021,VRBrunnstrom2020,VRZheleva2020}. This challenge becomes even more critical in the resource constrained mobile edge computing (MEC) scenarios, where the end-user experience is tightly coupled to the last-mile network conditions.

A key use case that exemplifies the importance of throughput prediction is adaptive video streaming, particularly Dynamic Adaptive Streaming over HTTP (DASH), which has become the industry standard for delivering video content over variable networks. DASH works by dividing video into small, time-segmented chunks encoded at multiple quality levels, allowing the client to dynamically select the most appropriate bitrate for each chunk based on the current network conditions. However, due to  unwarranted throughput fluctuations, reactive bitrate adaptation based on recent measurements may trigger rebuffering events, video stalls, or frequent quality shifts, each of which adversely affects user-perceived QoE. To mitigate this, DASH clients can benefit significantly from predictive throughput models that estimate near-future network performance rather than relying solely on past or current metrics. Such predictions must account for diverse and often hidden parameters, such as Reference Signal Received Power (RSRP), Received Signal Strength Indicator (RSSI), device mobility, user density, radio access technology (RAT), and base station load, which jointly determine the achievable transport layer throughput. Consequently, accurate, real-time throughput prediction becomes essential for stabilizing video quality, minimizing playback interruptions, and optimizing end-to-end streaming performance, particularly in 5G-enabled edge environments, which are characterized by user and network heterogeneity~\cite{tran2018adaptive,tuysuz2020qoe,farahani2022ararat}.

Despite the critical importance of throughput prediction in optimizing adaptive streaming and other bandwidth-sensitive services, achieving accurate and scalable prediction in 5G/6G networks remains a non-trivial task. The predictive performance of existing approaches is significantly affected by several real-world complexities that arise in operational mobile edge environments. These complexities stem from the highly heterogeneous nature of devices and networks, the non-IID nature of the data across different users and locations, and the trade-offs between personalization, generalization, and privacy preservation. Addressing these challenges requires a rethinking of conventional supervised learning paradigms that assume centralized data access and homogeneous training conditions.

\textbf{Challenge 1: \textit{Device and Network Heterogeneity.}}  
Modern 5G deployments are inherently heterogeneous, supporting a wide spectrum of user equipment (UE), including smartphones, IoT sensors, connected vehicles, and UAVs, operating under varying mobility, power constraints, and access technologies. In addition, these devices connect to networks that differ by carrier, radio access technology (e.g., 4G, 5G SA/NSA), frequency bands, and base station densities. Each of these factors introduces distinct throughput characteristics. A model trained on one combination of device and network type often fails to generalize across others, limiting its real-world applicability. Therefore, throughput prediction models must account for this diversity by adapting to different operational contexts, rather than relying on static or one-size-fits-all training pipelines.

\textbf{Challenge 2: \textit{Non-IID and Distributed Data.}}  
The distribution of network traces and throughput characteristics across users is often not independent and identically distributed (non-IID), due to factors such as geographic locations, user behavior, network provisioning, and mobility patterns. Most state-of-the-art ML and DL-based throughput prediction approaches~\cite{Yue2018, Raca2019, Narayanan2021, Elsherbiny2020, Palit2023, biswas2023kalman, Sen2023} are trained using centralized datasets collected under controlled settings. Such datasets fail to capture the real-world statistical heterogeneity present in operational networks. As a result, models trained in these environments may show good offline performance but degrade sharply  in diverse and decentralized conditions. Moreover, privacy regulations and bandwidth limitations often restrict the central collection of throughput traces, further motivating the need for distributed learning paradigms respecting data locality.

\textbf{Challenge 3: \textit{Balancing Model Generalization with Personalization.}}  
To be practically useful, a throughput predictor must generalize well across unseen network conditions while also offering personalized adaptation to local context, e.g., a user-specific device, location, or connectivity history. Achieving this dual goal is particularly challenging in distributed learning scenarios, such as in federated learning (FL), where data remains decentralized and models are trained collaboratively without sharing raw data. The inherent heterogeneity in client datasets, both in size and distribution, can lead to skewed updates and convergence issues in FL training. At the same time, naive aggregation strategies may overfit to dominant patterns, reducing prediction quality for minority users or edge cases. Designing throughput predictors that can jointly preserve generalization, enable personalization, and operate efficiently under FL constraints remains a core research challenge.

Throughput prediction has been extensively studied using a broad range of methodologies, including statistical time-series models such as \ac{ARIMA}, exponential smoothing, and moving averages~\cite{Makridakis2018, adhikari2013introductory, Yin2015}, as well as supervised machine learning and deep neural networks~\cite{Zhang2019, Raca2020_1, Schmid2021,Mondal2020,Palit2023,biswas2023kalman,Yue2018}. While these approaches have shown promise in controlled environments, they largely rely on centralized data collection and training pipelines, which limit their ability to capture the diverse and dynamic nature of real-world mobile edge deployments. These models often assume uniform device capabilities and network conditions, thereby failing to accommodate the heterogeneity in user equipment, radio access technologies, and mobility patterns that influence throughput in practice. Moreover, they typically operate under the assumption of identically distributed data across clients, which does not hold in federated scenarios where client datasets are inherently non-IID. As a result, such models struggle to generalize effectively across distributed settings and cannot offer adequate personalization without compromising on privacy. Although our recent efforts, such as FedPut~\cite{Sen2023}, have explored FL for throughput prediction, their scope remains limited to specific model types (e.g., LSTM) and aggregation strategies (e.g., \textit{FedAvg}), without examining how different combinations of models and federated algorithms perform under heterogeneous system and data conditions. Consequently, existing literature provides only a partial view of the design space and lacks a systematic framework for benchmarking federated throughput predictors that are robust, scalable, and well-suited to the requirements of real-time adaptive applications.

In this work, we address the limitations of existing throughput prediction methods by developing a comprehensive benchmarking framework that systematically evaluates FL strategies for mobile edge environments. Our approach is explicitly designed to tackle the three key challenges outlined earlier. First, to address device and network heterogeneity, we benchmark multiple model architectures, ranging from shallow convolutional models to deep temporal predictors like LSTMs and Transformers, on diverse 5G datasets that capture a variety of devices, mobility patterns, and radio configurations. Second, to handle the non-IID and distributed nature of real-world data, we assess the performance of different FL aggregation strategies, including \textit{FedAvg}~\cite{Mcmahan2016}, \textit{FedProx}~\cite{Sahu2020}, and \textit{FedBN}~\cite{Dou2021}, each of which offers varying levels of robustness to client heterogeneity. Finally, to balance model generalization with personalization, we examine how combinations of architectures and FL strategies affect convergence, accuracy, and variance across clients, and integrate the most effective configurations into a proof-of-concept adaptive live-streaming system. This end-to-end evaluation allows us to quantify the practical utility of federated throughput prediction in enhancing real-time QoE. In contrast to the existing works, our contributions in this paper are as follows. 

\textbf{Contribution 1: \textit{First systematic benchmarking of model–strategy combinations for federated throughput prediction.}}  
To the best of our knowledge, this is the first work to systematically benchmark a wide range of FL aggregation strategies, \textit{FedAvg}~\cite{Mcmahan2016}, \textit{FedProx}~\cite{Sahu2020}, and \textit{FedBN}~\cite{Dou2021}, across multiple time-series model architectures including CNNs, LSTMs, {LSTM}+{CNN} hybrids, and Transformers~\cite{vaswani2017}, for the task of throughput prediction. Our study reveals previously unexplored interactions between architecture and aggregation method under varying levels of client heterogeneity, offering concrete guidance on choosing suitable configurations based on convergence speed, accuracy, and stability.

\textbf{Contribution 2: \textit{First empirical evaluation of FL robustness under real-world 5G heterogeneity.}}  
Unlike prior works that rely on synthetic or homogenous datasets, our evaluation spans five realistic and diverse 5G throughput datasets~\cite{Narayanan2020, Narayanan2021,Raca2020_1}, reflecting real-world variations in device type, network deployment, mobility, and geographical coverage. We quantify the impact of non-IID data, cohort composition, and client diversity on FL performance, and identify \textit{FedBN} as a robust choice for non-IID settings, highlighting its practical relevance for edge deployments. This is the first study to provide such a fine-grained empirical characterization of heterogeneity in the context of federated throughput forecasting.

\textbf{Contribution 3: \textit{First integration of FL-based throughput prediction into a real-time QoE optimization loop.}}  
We go beyond predictive accuracy and demonstrate the end-to-end utility of federated models in an operational context. By integrating our best-performing FL models into a DASH-compliant live video streaming system~\cite{LSQoESun2019}, we show that accurate, decentralized throughput forecasting can meaningfully improve user-perceived QoE, reducing stall times, lowering latency, and smoothing bitrate transitions. This is the first work to operationalize federated throughput prediction in a practical media delivery pipeline and evaluate its real-world impact on adaptive streaming.

\begin{table*}[!t]
     \scriptsize
     \centering
     \caption{State-of-the-art (SOTA) approaches for throughput prediction in comparison with the proposed method.}
     \begin{tabular}{|p{0.3cm}|c|p{1.5cm}|p{4cm}|p{5cm}|p{1.1cm}|p{1.3cm}|}
     \hline
     \textbf{Sl. No:} & \textbf{Ref.} & \textbf{Network (3G/4G/5G or WiFi)} & \textbf{Method used} & \textbf{Geographical Area Covered} & \textbf{Centralized/ Distributed} & \textbf{Application} \\ \hline
     1 & Khan et. al.~\cite{KHAN2020} & WiFi & {MLP}, {SVR}, {RF}, {DT} & Real home WiFi network. & Centralized & - \\ \hline
     2 & Yue et.al.~\cite{Yue2018} & 4G LTE & {RF} & Indoor and Outdoor 4G cellular data, collected in Kingston Ontario. & Centralized & - \\ \hline
     3 &  Raca et. al.~\cite{Raca2019} & 4G LTE & {RF} & Static, pedestrian, bus, train, car, and highway. & Centralized & \ac{ABR} \ac{VoD} streaming \\ \hline
     4 & Kousias et. al.~\cite{Kousias2019} & 4G LTE & {RF}, {MLR}, and {SVR} & Driving datasets in Austria and Norway. & Centralized & - \\ \hline
     5& Mondal et.al.~\cite{Mondal2020} & 4G/LTE & RF & Metropolitan and sub-urban cities of India. & Centralized & \ac{ABR} \ac{VoD} streaming\\ \hline
       6 & Hameed et.al.~\cite{Hameed2021} & IoT & {DT}, {GBDT}, {SVR}, \ac{KNN}, {MLR} & Smart buildings and traffic from different applications like HVAC, Lighting, Emergency, Surveillance, \ac{AR}, VoIP. & Centralized & -\\ \hline
       7 & Schmid et. al.~\cite{Schmid2019} & 4G LTE & {RF}, {SVR}, {LSTM},  \ac{FFNN} & Collected from two cities Amberg and Aschaffenburg. & Centralized & - \\ \hline  
     8 & Zhang et.al.~\cite{Chauting2019} & GSM, CDMA, 4G LTE & Transfer Learning (TL) based \ac{LSTM} & Across different regions in the city of Milan. & Centralized & -\\ \hline
     9 & Mei et.al.~\cite{Mei2020} &  4G LTE and HSPA & \ac{LSTM} RNN &  Long bandwidth traces on New York City MTA bus and subway. & Centralized & -\\ \hline
     10 & Raca et.al.~\cite{Raca2020} & 4G LTE & {LSTM}, {RF}, {SVR}, & Indoor and Outdoor 4G cellular data, collected in Kingston Ontario~\cite{Yue2018}. & Centralized & \ac{ABR} \ac{VoD} streaming \\ \hline
    11 & Minovski et.al.~\cite{Minovski2021} & 4G LTE, 5G SA, and NSA  &  {RF},  XGBoost, SVR, {MLP} & Driving in urban, suburban, and rural areas, as well as tests in large crowded areas. & Centralized & - \\ \hline
    12 & Narayanan et.al.~\cite{Narayanan2020} & 5G & DL based gradient boosting and sequence-to-sequence algorithms &  Urban areas covering roads, railroad crossings, restaurants, coffee shops, and outdoor recreational parks in Minneapolis. & Centralized &  \\ \hline
    13 & Elsherbiny et.al.~\cite{elsherbiny20204g} & 4G LTE &  {ARIMA}, {KNN}, {SVR}, Ridge Regression, {RF}, {LSTM}  & Indoor and Outdoor 4G cellular data, collected in Kingston Ontario~\cite{Yue2018}. & Centralized & - \\ \hline
    14 & Qu et.al~\cite{Jiantao2020} & LTE Railways & {CNN} + Bi-{LSTM} with attention layer, Evolutionary attention based {LSTM}, {LSTM}+CNN, Recurrent Neural Network & Wide Geographical region, Urban, suburban, and rural, connected by Shuohuang railway. & Centralized & - \\ \hline 
    15 & Narayanan et.al.~\cite{Narayanan2021} & 4G LTE, 5G SA and NSA & Harmonic Mean (HM), {GBDT} & Urban areas covering roads, railroad crossings, restaurants, coffee shops, and outdoor recreational parks in Minneapolis~\cite{Narayanan2020}. & Centralized & \ac{ABR} \ac{VoD} Streaming \\ \hline
    16 & Palit et.al~\cite{Palit2023} & 4G LTE, 5G SA and NSA & {LSTM} based Transfer learning & 4G data collected in different cities in India~\cite{Mondal2020}, Urban areas of Minneapolis~\cite{Narayanan2020, Narayanan2021}, Urban areas of Ireland~\cite{Raca2020_1}, A synthetic 5G dataset. & Centralized & ABR \ac{VoD} streaming \\ \hline
    17 & Biswas et.al.~\cite{biswas2023kalman} & 5G SA and NSA & Kalman filter based {MLR}, {ARIMA}, {RF}, SVM, {XGBoost} {LSTM} &  Urban areas of Minneapolis~\cite{Narayanan2020, Narayanan2021}, Urban areas of Ireland~\cite{Raca2020_1}. & Centralized & \ac{ABR} \ac{VoD} streaming, Live-streaming  \\ \hline
    18 & Sen et.al.~\cite{Sen2023} & 4G, 5G SA and NSA & {LSTM} based FL with \fedavg{}, {ARIMA}, {RF}, {LSTM} & 4G data collected in different cities in India~\cite{Mondal2020}, urban areas of Minneapolis~\cite{Narayanan2020, Narayanan2021}, urban areas of Ireland~\cite{Raca2020_1}, A synthetic 5G dataset. & Federated & \ac{ABR} \ac{VoD} streaming \\ \hline
    \rowcolor{gray!30}19 & \textbf{Our Work} & 5G SA  and NSA & Federated learning with different combinations of model architectures, such as \ac{CNN}, \ac{LSTM}, \ac{LSTM}+\ac{CNN}, and Transformers, and different federated averaging strategies, such as \fedavg\ , \fedprox\ , \fedbn\  & Urban areas of Minneapolis~\cite{Narayanan2020, Narayanan2021}, Urban areas of Ireland~\cite{Raca2020_1}. & Federated & Live Streaming\\ \hline
    \end{tabular}
    \label{tab:SOTA}
\end{table*}
\begin{table*}
 \centering
 \caption{{Details of the throughput datasets used in this work.}}\label{tab:dataset_det}
\begin{tabular}{|c|c|c|c|c|c|}
\hline
\textbf{Dataset} & \textbf{Network Service Provider and Type} & \textbf{Collected in} & \textbf{Smartphone Models} & \textbf{Mobility} & \textbf{Application}\\ \hline
\textbf{MNWild-Ver~\cite{Narayanan2021}}      & Verizon, 5G Default                        & Minneapolis           & Samsung Galaxy S20 Ultra 5G  & Walking         & File Download \\ \hline
\textbf{MMNWild-TNSA~\cite{Narayanan2021}}    & TMobile, 5G Non-Stand Alone                & Minneapolis           & Samsung Galaxy S20 Ultra 5G  & Walking     & File Download     \\ \hline
\textbf{MNWild-TSA~\cite{Narayanan2021}}      & TMobile, 5G Stand Alone                    & Minneapolis           & Samsung Galaxy S20 Ultra 5G  & Walking      & File Download    \\ \hline
\textbf{LUMOS-5G~\cite{Narayanan2020}}         & Verizon, 5G Non-Stand Alone                & Minneapolis           & Samsung Galaxy S10                   & Driving      & Video Streaming   \\ \hline
\textbf{Irish~\cite{Raca2020_1}}         & An Irish Mobile Network Provider, 5G       & Unspecified Irish city       & Samsung Galaxy S10                   & Static      & File Download     \\ \hline
\end{tabular}
\end{table*}

\section{Background and Related Work}
\label{sec:relwork}
Accurate prediction of network throughput has long been a critical requirement for enabling high-quality network-aware applications such as adaptive video streaming~\cite{Yin2015}. We start with a discussion of the  background of the Federated learning (FL) architecture and a brief survey of the existing throughput prediction mechanisms proposed in the literature.

\subsection{Federated Learning}\label{sec:FL}
Federated learning (FL) is a decentralized machine learning paradigm where a central server coordinates training across a set of distributed client devices, each possessing private data. Instead of transmitting raw data, clients compute model updates locally and send them to the server, thereby preserving privacy and reducing communication overhead~\cite{mcmahan2017communication}. This makes FL well-suited for 5G edge computing scenarios, where data is inherently decentralized and collected across a heterogeneous device landscape.

In the canonical Federated Averaging (\fedavg{}) algorithm~\cite{kairouz2021advances}, each client \( k \in \{1, \dots, K\} \) trains a local model \( w_k^t \) on its dataset \( D_k \) during round \( t \), and the server updates the global model via the weighted aggregation: 
\[
w^{t+1} = \sum_{k=1}^K \frac{n_k}{n} w_k^t,
\]
where \( n_k = |D_k| \) and \( n = \sum_{k=1}^K n_k \). While \fedavg{} is simple and effective in IID settings, its performance can degrade in heterogeneous environments due to divergence among client models~\cite{Sahu2020}. To address this, several advanced strategies have been developed to improve robustness.

\subsection{Alternate Federated Aggregation Strategies}

\textbf{\textit{FedProx}}~\cite{Sahu2020} enhances \fedavg{} by adding a proximal term to the client loss function, which penalizes any deviation from the global model. Each client solves the following objective during local training:
\begin{equation}
    \min_{w} \, f_k(w) + \frac{\mu}{2} \| w - w^t \|^2,
\end{equation}
where \( \mu \) is a tunable regularization parameter. This constraint mitigates model divergence, making \textit{FedProx}  useful in the presence of system-level and data-level heterogeneity.

\textbf{\textit{FedBN}}~\cite{Dou2021} addresses statistical heterogeneity by decoupling batch normalization (BN) layers during model aggregation. Instead of averaging BN statistics globally, each client maintains its own BN parameters while synchronizing the remaining weights. This personalization strategy is  effective in non-IID environments, as it allows models to adapt to local data distributions while benefiting from shared representations.

Both \textit{FedProx} and \textit{FedBN} aim to overcome the limitations of \fedavg{} in real-world federated settings, enabling faster convergence and improved generalization in the presence of heterogeneous data. These techniques form the foundation for our benchmarking framework, which evaluates their impact in the context of 5G throughput prediction and adaptive live video streaming.

\subsection{Network Throughput Prediction}
Throughput prediction is a time-series forecasting task influenced by several complex and interacting parameters from multiple layers of the network stack~\cite{Yue2018,Raca2018_2,Sen2023,biswas2023kalman}. Traditional statistical methods such as \ac{MA}~\cite{adhikari2013introductory}, \ac{ARMA}, \ac{ARIMA}~\cite{elsherbiny20204g}, harmonic mean~\cite{Yin2015}, and \ac{EWMA}~\cite{adhikari2013introductory} have been explored, but often suffer from limited flexibility in capturing nonlinear dependencies and adapting to dynamic network environments. Table~\ref{tab:SOTA} summarizes a wide range of centralized approaches proposed for throughput prediction across various access technologies.

Linear models such as \ac{MLR} have been widely used in early works for bandwidth prediction over {WCDMA} networks~\cite{Nasri2019},  and diverse {IoT} settings like HVAC and lighting systems~\cite{Hameed2021}.   While lightweight and interpretable, these models often fail to capture complex relationships in modern cellular networks. For improved performance, tree-based methods like \ac{RF} have been introduced~\cite{Yue2018,Raca2019,Kousias2019,Mondal2020,Schmid2019,Raca2020}, demonstrating better accuracy compared to MLR and \ac{SVR}~\cite{Kousias2019}. Deep learning methods, such as \ac{LSTM}, have also been adopted for their ability to model long-term temporal dependencies in 4G traces~\cite{Chauting2019,Mei2020,Raca2020,elsherbiny20204g,Jiantao2020,Palit2023}. WiFi throughput prediction has also been tackled using a variety of ML methods including \ac{MLP}, \ac{DT}, and \ac{SVR}~\cite{KHAN2020}.

As 5G networks have become more prominent, recent works~\cite{Narayanan2020,Narayanan2021,Palit2023,biswas2023kalman} have begun to address the unique challenges posed by throughput prediction in these settings. For example,~\cite{Narayanan2020} and~\cite{Narayanan2021} model 5G throughput using both regression and classification techniques with \ac{GBDT} and sequence-to-sequence architectures. These models rely on comprehensive traces collected from mobile networks across various mobility conditions and devices. In~\cite{biswas2023kalman}, Kalman filters are applied over baseline \ac{MLR} models to further refine predictions. However, despite these improvements, all of the above models are still centralized, requiring clients to share their raw data with a central server. This setup not only violates user privacy but also assumes uniformity across network environments, which is an assumption that is increasingly invalid in 5G deployments.

\subsection{Limitations of Existing Works}

As we observe from the above discussion, several approaches have been proposed in the literature to address the throughput prediction problem, ranging from statistical forecasting models~\cite{adhikari2013introductory,elsherbiny20204g} to advanced ML and DL architectures~\cite{Mei2020,Raca2020,Sen2023}. However, most existing solutions rely on centralized learning paradigms where a global model is trained using data aggregated from all clients or user devices~\cite{Narayanan2020,biswas2023kalman}. This assumption creates multiple limitations, particularly in the context of 5G networks, which are inherently heterogeneous in both device capabilities and network environments~\cite{Narayanan2021}. In such scenarios, centralized approaches often fail to generalize across diverse deployment conditions, raising concerns around data privacy and scalability~\cite{mcmahan2017communication}.

Recognizing this limitation,~\cite{Sen2023} proposed FedPut, a federated LSTM-based approach using \fedavg{} to enable decentralized throughput prediction. Although FedPut marks a significant step toward privacy-preserving prediction, it is limited in scope- it evaluates only one aggregation strategy and one model architecture, leaving open questions around the effectiveness of alternate FL configurations.

In contrast, our work addresses these gaps by developing a comprehensive benchmarking framework using the \textbf{Flower} FL framework and \textbf{Ray} for scalable client simulation. We systematically evaluate combinations of model architectures (CNN, \ac{LSTM}+\ac{CNN}, LSTM,  and Transformers) with various FL aggregation strategies (\fedavg{}, \fedprox{}, and \fedbn{}) for throughput prediction in 5G settings. We also demonstrate how these models improve prediction accuracy and streaming QoE of live video streaming services. To the best of our knowledge, this is the first work that applies FL to enhance the QoE of live video streaming via throughput-aware bitrate selection.

\section{Exploratory Data Analysis}\label{sec:data_analysis}
Before detailing the design of our benchmarking pipeline, we first present an in-depth analysis of the publicly available 5G throughput datasets to establish their non-IID nature. The details follow. 

\subsection{Publicly Available 5G Datasets}\label{public_data}
We have used five publicly available 5G throughput datasets -- (1) \textbf{MNWild-Ver}~\cite{Narayanan2021}, (2) \textbf{MNWild-TNSA}~\cite{Narayanan2021}, (3) \textbf{MNWild-TSA}~\cite{Narayanan2021}, (4) \textbf{Lumos-5G}~\cite{Narayanan2020}, and (5) \textbf{Irish}~\cite{Raca2020_1}. The details of these datasets are provided in the Table~\ref{tab:dataset_det}, while the corresponding feature space in Table~\ref{tab:feature_space}.

\begin{table}[t]
    \scriptsize
    \centering
    \caption{Raw features available across different datasets}
    \label{tab:feature_space}
    \setlength{\tabcolsep}{1pt} 
    \renewcommand{\arraystretch}{1.1} 
    \begin{tabular}{|c|>{\centering\arraybackslash}p{0.9cm}|>{\centering\arraybackslash}p{0.9cm}|>{\centering\arraybackslash}p{0.9cm}|>{\centering\arraybackslash}p{0.9cm}|>{\centering\arraybackslash}p{0.9cm}|}
    \hline
        \textbf{Feature Name} 
        & \textbf{MNWild-Ver} 
        & \textbf{MNWild-TSA} 
        & \textbf{MNWild-TNSA} 
        & \textbf{Lumos-5G} 
        & \textbf{Irish} \\ \hline
        
        Timestamp & $\pmb{\checkmark}$ & $\pmb{\checkmark}$ & $\pmb{\checkmark}$ & $\pmb{\checkmark}$ & $\pmb{\checkmark}$ \\ \hline
        Lat, Long & $\pmb{\checkmark}$ & $\pmb{\checkmark}$ & $\pmb{\checkmark}$ & $\pmb{\checkmark}$ & $\pmb{\checkmark}$ \\ \hline
        Radio type (4G/5G) & $\pmb{\checkmark}$ & $\pmb{\checkmark}$ & $\pmb{\checkmark}$ & $\pmb{\checkmark}$ & $\pmb{\checkmark}$ \\ \hline
        Speed & $\pmb{\checkmark}$ & $\pmb{\checkmark}$ & $\pmb{\checkmark}$ & $\pmb{\checkmark}$ & $\pmb{\checkmark}$ \\ \hline
        Operator Name & $\pmb{\checkmark}$ & $\pmb{\checkmark}$ & $\pmb{\checkmark}$ & X & X \\ \hline
        Horizontal Handover & $\pmb{\checkmark}$ & $\pmb{\checkmark}$ & $\pmb{\checkmark}$ & $\pmb{\checkmark}$ & $\pmb{\checkmark}$ \\ \hline
        Vertical Handover & X & X & X & $\pmb{\checkmark}$ & X \\ \hline
        Signal Strength (RSRP/RSRQ/RSSI) & $\pmb{\checkmark}$ & $\pmb{\checkmark}$ & $\pmb{\checkmark}$ & $\pmb{\checkmark}$ & $\pmb{\checkmark}$ \\ \hline
        SNR & $\pmb{\checkmark}$ & $\pmb{\checkmark}$ & $\pmb{\checkmark}$ & $\pmb{\checkmark}$ & $\pmb{\checkmark}$ \\ \hline
        CQI & X & X & X & X & $\pmb{\checkmark}$ \\ \hline
        Throughput & $\pmb{\checkmark}$ & $\pmb{\checkmark}$ & $\pmb{\checkmark}$ & $\pmb{\checkmark}$ & $\pmb{\checkmark}$ \\ \hline
        Data State & X & X & X & X & \checkmark \\ \hline
    \end{tabular}
\end{table}
\begin{figure}[!ht]
    \centering
    \begin{subfigure}[t]{0.47\columnwidth}
        \centering
        \includegraphics[width=\linewidth, trim={1cm 9cm 1cm 9cm}]{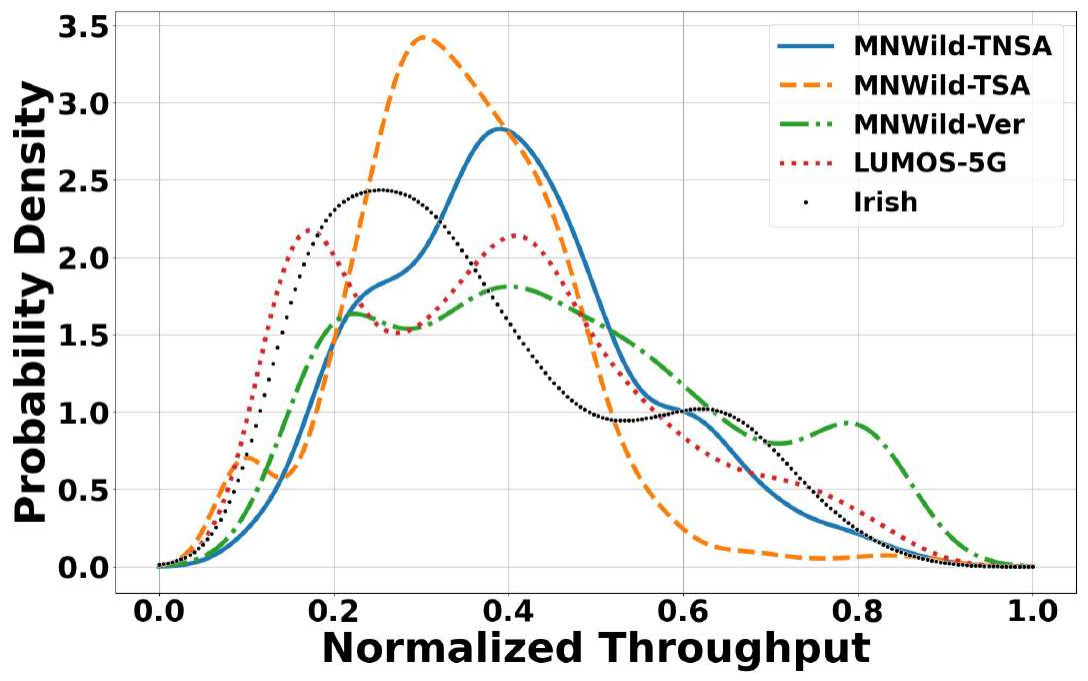}
        \caption{Kernel Density Estimates.}
        \label{fig:KDE}
    \end{subfigure}
    \hfill%
    \begin{subfigure}[t]{0.51\columnwidth}
        \centering
        \includegraphics[width=\linewidth,trim={1cm 9cm 4cm 9cm}]{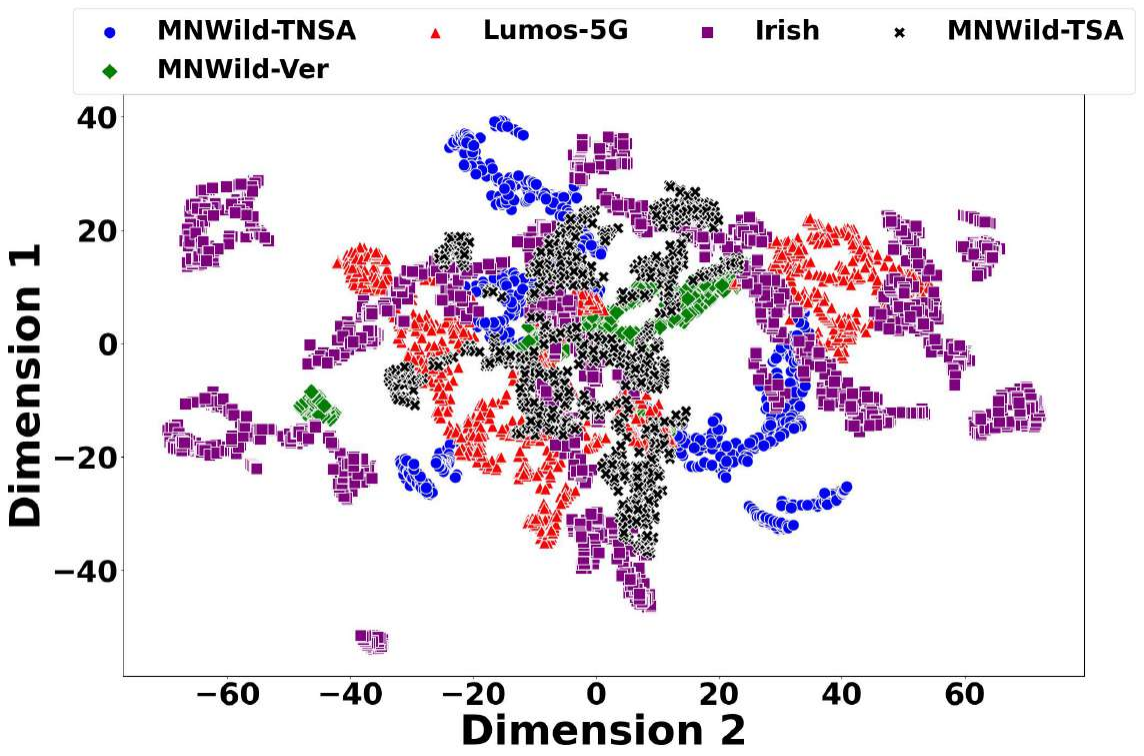}
        \caption{t-SNE embeddings.}
        \label{fig:tsne}
    \end{subfigure}
    \caption{Illustration of distributional heterogeneity across different datasets.}
    \label{fig:dist_plots}
\end{figure}

The throughput datasets used in this study have been collected from multiple cities under diverse controlled environments. These datasets span various mobile network service providers and deployment configurations, including 4G, and both standalone (SA) and non-standalone (NSA) modes of 5G mmWave technologies. Data collection was performed using commercially available smartphones from different manufacturers, covering heterogeneous hardware and radio capabilities. Furthermore, multiple data recording tools were used, including Ookla Speedtest\footnote{\url{https://www.speedtest.net/}, accessed on \today}~\cite{Narayanan2021} and GNetTrackPro\footnote{\url{https://www.gyokovsolutions.com/G-NetTrack\%20Android.html}, accessed on \today}~\cite{Raca2020_1}, under varying mobility patterns and workload conditions.

Notably, the recorded throughput exhibits observable variations arising from differences in device types, data collection software (e.g., browser-based vs. native applications), and network operator configurations. Fig.~\ref{fig:KDE} presents the kernel density estimation (KDE) of normalized throughput traces from four representative datasets using a Gaussian kernel with bandwidth 1. The resulting distributions show clear shifts across datasets, indicating differences in the underlying network behavior. Although the KDE curves have been normalized for comparison, the actual throughput ranges vary significantly due to disparities in radio access technology and deployment settings.

\figurename\ref{fig:tsne}, which visualizes the t-SNE embeddings of the same datasets, also corroborates the data heterogeneity. The visual separation of data clusters is indicative of localized, device-specific, operator-dependent patterns, which collectively contribute to the non-IID nature of the data. Such heterogeneity poses a fundamental challenge for global throughput prediction models trained under homogeneity assumptions, as they are likely to exhibit degraded performance when deployed across unseen client configurations. We, therefore, proceed with a detailed analysis of the feature space to better understand the underlying disparities and their implications for FL.

\begin{table}[t]
  \centering
  \small
  \caption{Pearson's correlation coefficient ($\rho$) of the present network parameters with future throughput for different Prediction Window length in seconds. MNWild-TNSA has similar correlation between parameters as MNWild-TSA (hence, not shown in the table).}
  \label{table:feature_correlation}
\begin{tabular}{|p{1.2cm}|p{1cm}|p{1cm}|p{1cm}|p{1cm}|p{1cm}|}
\hline
 &  & \multicolumn{4}{c}{\textbf{Dataset}}  \\ \cline{3-6}
 \multirow{2}{*}{\textbf{Features}} & \multirow{2}{*}{\parbox{1cm}{\textbf{Window\\ Size}}} & \textbf{MNWild-Ver} & \textbf{MNWild-TSA} & \textbf{Lumos-5G} & \textbf{Irish}  \\ \hline
\hline
\multirow{3}{*}{\textbf{Latitude}} &  1s & 0.352 & -0.095 & 0.339 & -0.117 \\ \cline{2-6} 
 &  3s & 0.358 & -0.104 & 0.333 & -0.116 \\ \cline{2-6} 
 &  5s & 0.364 & -0.113 & 0.325 & -0.116 \\ \hline\hline
\multirow{3}{*}{\textbf{Longitude}} &  1s & -0.042 & -0.201 & -0.295 & -0.130 \\ \cline{2-6} 
 &  3s & -0.056 & -0.198 & -0.284 & -0.130 \\ \cline{2-6} 
 &  5s & -0.070 & -0.195 & -0.270 & -0.130 \\ \hline\hline
\multirow{3}{*}{\textbf{Speed}} &  1s & -0.164 & 0.105 & -0.417 & -0.107 \\ \cline{2-6} 
 &  3s & -0.179 & 0.113 & -0.446 & -0.111 \\ \cline{2-6} 
 &  5s & -0.190 & 0.113 & -0.459 & -0.117 \\ \hline\hline
\multirow{3}{*}{\textbf{RSRP}} &  1s & 0.140 & 0.506 & 0.436 & 0.245 \\ \cline{2-6} 
 &  3s & 0.138 & 0.467 & 0.401 & 0.241 \\ \cline{2-6} 
 &  5s & 0.133 & 0.428 & 0.376 & 0.239 \\ \hline\hline
\multirow{3}{*}{\textbf{SINR}} &  1s & 0.690 & 0.024 & 0.365 & 0.149 \\ \cline{2-6} 
 &  3s & 0.689 & 0.020 & 0.340 & 0.149 \\ \cline{2-6} 
 &  5s & 0.675 & 0.018 & 0.324 & 0.149 \\ \hline\hline
\multirow{3}{*}{\textbf{Through-}} &  1s & 0.983 & 0.972 & 0.974 & 0.945 \\ \cline{2-6} 
 &  3s & 0.920 & 0.867 & 0.853 & 0.647 \\ \cline{2-6} 
{\textbf{put}} &  5s & 0.879 & 0.803 & 0.768 & 0.369 \\ \hline
\end{tabular}
\end{table}

\subsection{Feature Analysis}
To further characterize the inherent heterogeneity of the datasets, we analyze the correlation between current network parameters and future throughput. Table~\ref{tab:feature_space} summarizes the set of features consistently available across all datasets considered in this study. Our primary objective is to predict the future downlink throughput based on currently observed network conditions. Accordingly, Table~\ref{table:feature_correlation} presents the Pearson correlation coefficients between each feature and the future throughput, evaluated over prediction horizons of $F = 1$, $3$, and $5$ seconds. From the results, it is evident that the current throughput, RSRP, and Signal-to-Interference-plus-Noise Ratio (SINR) exhibit the strongest correlations with future throughput across all datasets and time windows. As expected, the predictive correlation generally declines with the increasing forecast horizon $F$, reflecting reduced temporal dependency at longer intervals.

It may be observed that the magnitude and sign of the correlation coefficients vary across datasets, implying that the influence of individual features on future throughput is dataset-specific. This variation emphasizes the statistical heterogeneity of the data and highlights the limitations of a one-size-fits-all prediction model, thereby reinforcing the necessity of adopting a distributed and personalized learning paradigm, such as FL, to accommodate client-specific data distributions.

\section{Methodology}\label{sec:methodlogy}
This section presents our benchmarking framework for evaluating FL-based throughput prediction in heterogeneous 5G environments. The proposed architecture is built over our initial work \emph{FedPut}~\cite{Sen2023};  we extend the solution by incorporating a wider range of client cohorts, aggregation strategies, and model architectures. This section outlines the system architecture and key components used in the  study.

\subsection{Problem Statement}
Say a mobile device $k$ from a geographical area $\mathcal{G}$ is connected to an available operator $\mathcal{X}$ running an application $\mathcal{A}$. The primary objective of the federated throughput prediction algorithm in this work is to develop a \textbf{robust} \textit{in-situ} model that can seamlessly predict the 5G cellular network throughput $\hat{\tau}_k(n+F)$ from the current and historical throughput traces and network parameters, as sensed and estimated by the mobile device. Mathematically, 
\begin{multline}\label{eq:timeseries}
   \hat{\tau}_k(n+1:n+F) = f(\tau_k(n),\tau_k(n-1),\cdots,\tau_k(n-H),\\ \mathbf{x}_k(n),\mathbf{x}_k(n-1),\cdots, \mathbf{x}_k(n-H)),  
\end{multline}
where $H$ and $F$ denote the history and the prediction window lengths, respectively. $\mathbf{x}(n)$ represents the feature vector of the network parameters and $\tau_k(n)$ represents the throughput at the time step $n$. Here, robustness means that the framework should be resilient to the fluctuations caused by the hardware components of $\mathcal{M}$, the area-specific characteristics like population density, and the operator-specific network variations. 

Given the heterogeneity of 5G deployments, a personalized \textbf{in-situ learning} can potentially capture the device-specific and network-specific parameters. However, such methods demand an extensive training dataset that is difficult to obtain due to the nascent deployment of 5G primarily in the middle and low-economy countries. So, in our prior work~\cite{Sen2023}, we hypothesized that an \emph{FL}-based approach can mitigate this problem of dearth in the training datasets and still train a robust model by extracting knowledge collaboratively across different devices~\cite{Konecny2016, konevcny2015federated, Jakub2016}. However, Fedput~\cite{Sen2023} is limited in terms of its ability to capture the diversity in data across different devices and services. Understanding the challenges and opportunities presented by the datasets as identified in the pilot experiments and picking up from where~\cite{Sen2023} left, in this work, we design the benchmarking setup as follows.

\subsection{System Architecture}
\label{subsec:architecture}
   
\begin{figure}[t]%
	\centering
        \includegraphics[width=0.6
        \textwidth, trim = {6.2cm 7cm 5.7cm 5cm}
        ]{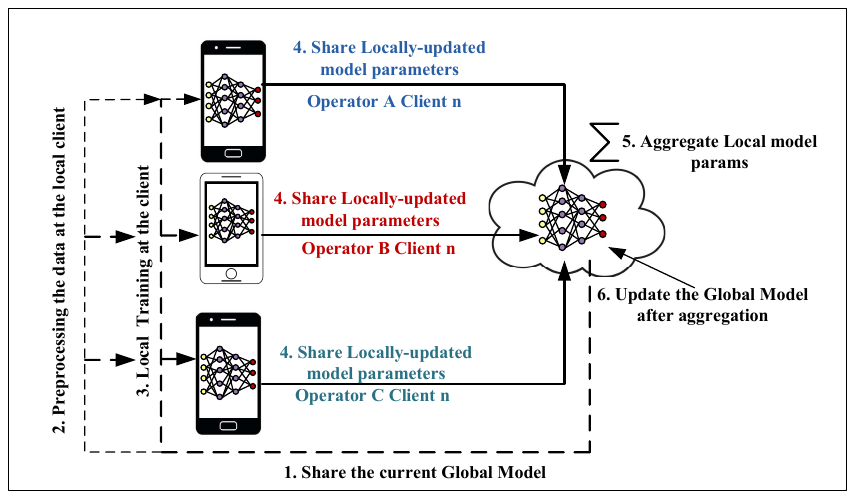}
	\caption{\newadd{\textbf{\textit{A brief overview of the overall setup.}} We consider a cross-device FL framework, with mobile devices as clients connected to heterogeneous 5G networks. A centralized server coordinates the FL process, including client selection, aggregation, and synchronization. Local training is performed independently on each client using device-specific data. The updated model parameters are transmitted to the server, which aggregates these parameters to update the global model and redistributes it to the clients for the next training round. }}
	\label{fig:framework}
\end{figure}
We adopt a standard cross-device FL setup for benchmarking throughput prediction models in mobile 5G environments, as illustrated in \figurename~\ref{fig:framework}. The setup consists of a centralized server and multiple mobile devices acting as federated clients. The server maintains a global model, which is periodically updated through iterative rounds of training and aggregation. Each client hosts a local copy of the model and performs training using device-specific throughput traces. To preserve data privacy, clients do not share raw data; instead, they transmit their locally updated model parameters to the server. The server aggregates these updates using a specified aggregation strategy and broadcasts the updated global model to all clients. This process continues for multiple communication rounds until convergence.
\subsubsection{Preprocessing}
\label{sec:preprocessing}
In typical 5G networks, several hidden parameters directly impact the throughput. However, most of these hidden factors cannot be measured on the user side. For example, the load condition of the base stations cannot be measured at the user end. Similarly, the users cannot quantify the effect of resource scheduling algorithms without input from the service providers. In this work, we treat the effect of these latent parameters as noise and remove it using a moving average filter to remove the effect of noise. The filter window size is a parameter that can be varied to obtain different results. In this work, we have chosen a filter window of $W=3$ seconds, as a higher filter window size implies a larger \ac{SNR}, and a lower window size becomes restrictive. 

Furthermore, to maintain uniformity across all the clients, we adopt min-max scaling and standard scaling individually for each dataset. This step ensures that all the features lie between zero and one, overriding any chance of one data presiding over the others. 

After preprocessing, we arrange each client's data in timesteps, as shown in (\ref{eq:timeseries}), to leverage its time-series nature for designing the model discussed in the following subsection. We format the data as follows. At every time step $n$,  we create a feature vector $\mathbf{x}_k(n)$ for client $k$, such that  $\mathbf{x}_k(n) = [x^{(1)}_k(n), x^{(2)}_k(n), ..., x^{(|\mathcal{F}|)}_k(n)]^T$, where $\mathcal{F}$~\footnote{It is to be noted that $\mathcal{F}$ denotes the set of input features, while $F$ denotes the prediction horizon length.} corresponds to the set of input features in the filtered dataset, and $\tau_k(n)$ represents the target throughput.  For a historical window size $H$, we create a 2D-matrix of $\vec{\psi}_{\mathbf{x}_k}(n)$ to represent the matrix of network parameters and location-related input features from timestep $n-H$ to $n$ timesteps, i.e., 
\begin{multline}\label{eq:createsample}
	{\vec{\psi}_{\mathbf{x}_k}(n)} \\= \begin{pmatrix}
		x^{(1)}_k(n-H) & x^{(1)}_k(n-H+1) & ... & x^{(1)}_k(n)\\
		x^{(2)}_k(n-H) & x^{(2)}_k(n-H+1) & ... & x^{(2)}_k(n)\\
		\vdots & \vdots & \vdots & \vdots\\
		x^{(|\mathcal{F}|)}_k(n-H) & x^{(|\mathcal{F}|)}_k(n-H+1) & ... & x^{(|\mathcal{F}|)}_k(n) \end{pmatrix}.
\end{multline}
The vector of downlink throughput from $n-H$ to $n$ timesteps is represented as., 
\begin{equation}\label{eq:createsample_thpt}
	\vec{\psi}_{\tau_k}(n) = [\tau_k(n-H),\tau_k(n-H+1),\cdots,\tau_k(n)]^T,
\end{equation}

\subsubsection{\textbf{Client-Side Local Training}}
\label{subsec:local-training}

\newadd{At each communication round, participating clients receive the latest global model from the server and perform local training using their private datasets. The input to each model is the preprocessed time-series matrix \(\vec{\psi}_{\mathbf{x}_k}(n) \in \mathbb{R}^{|\mathcal{F}| \times (H+1)}\), as defined in Section~\ref{sec:preprocessing}, representing a historical window of network and location-specific features. The corresponding target is the downlink throughput sequence \(\vec{\psi}_{\tau_k}(n) \in \mathbb{R}^{H+1}\).}

\newadd{For models such as Transformers, which require inputs as a sequence of feature vectors, the input matrix is transposed into
\(\vec{\psi}_{\mathbf{x}_k}(n)^T \in \mathbb{R}^{(H+1) \times |\mathcal{F}|}\),
where each token \(\mathbf{x}_k(n)\) encodes the state at time \(n\). Positional encodings are added to \(\mathbf{x}_k(n)\) to preserve temporal structure. The resulting sequence is processed through self-attention layers to capture short and long-range dependencies across the historical window.}
\begin{table*}[ht]
\scriptsize
\centering
\caption{Model architectures and training configurations.}
\label{tab:model-config}
\begin{tabular}{|p{1.5cm}|p{6cm}|p{1.5cm}|p{1cm}|p{1cm}|}
\hline
\textbf{Model} & \textbf{Description} & \textbf{Activation Function} & \textbf{Learning Rate} & \textbf{Batch Size} \\
\hline
\textbf{CNN} & Captures short-term dependencies and local patterns using 2D convolutional layers. Includes Leaky ReLU activation. & Leaky ReLU & 0.001 & 32 \\
\hline
\textbf{LSTM} & Models long-term temporal dependencies using LSTM units; mitigates vanishing gradient problems. & ReLU & 0.0003 & 32 \\
\hline
\textbf{LSTM+CNN} & Hybrid model where LSTM first captures temporal patterns, followed by CNN for feature refinement. & ReLU & 0.003 & 32 \\
\hline
\textbf{Transformer} & Utilizes self-attention and multi-head mechanisms to capture both short- and long-term dependencies in the sequence. & ReLU & 0.001 & 32 \\
\hline
\end{tabular}
\end{table*}

\newadd{Each client locally optimizes a supervised forecasting objective to predict the future throughput over a window  \(F\), s.t:
\[
\hat{\tau}_k(n+1:n+F) = f_{\theta_k} \left( \vec{\psi}_{\mathbf{x}_k}(n)^T,\vec{\psi}_{\mathbf{\tau}_k}(n) \right),
\]
where \(f_{\theta_k}(\cdot)\) is the model parameterized by \(\theta_k\), and \(\hat{\tau}_k(n+1:n+F) \in \mathbb{R}^F\) denotes the predicted throughput values. After the prediction at time-step $n$ is over, the model moves the historical window by $F$ time steps. The model is trained using the mean squared error (MSE) loss:
\[
\mathcal{L}_k = \frac{1}{F} \sum_{i=1}^{F} \left( \tau_k(n+i) - \hat{\tau}_k(n+i) \right)^2,
\]
where \(\tau_k(n+i)\) is the ground-truth throughput.} Each client trains its model for a fixed number of local epochs and transmits only the updated parameters \(\theta_k\) back to the server. No raw data is shared, ensuring local data privacy.

\newadd{This benchmarking study includes a variety of model architectures- CNN, LSTM+CNN, LSTM~\cite{Sen2023},  and Transformers- trained independently on each client, with different federated aggregation strategies, as discussed next.}

\subsubsection{\textbf{Server-Side Aggregation}}
\label{subsec:aggregation}
After each round of local training, the central server collects the model updates \(\{\theta_k\}\) from the participating clients and aggregates them to obtain the updated global model \(\theta^{(t+1)}\). As mentioned earlier, in this paper, in addition to using \fedavg{}~\cite{mcmahan2017communication} which is a commonly used baseline for aggregation,  we have also explored other strategies such as \fedprox{}~\cite{Sahu2020} and \fedbn{}~\cite{Dou2021} to address the non-IID  data distribution of the clients. These strategies are benchmarked across multiple model architectures to evaluate their efficacy under heterogeneous mobile network settings.


\section{Implementation Details}\label{sec:implementation}
\begin{figure*}[h]
\centering 
     \begin{subfigure}{0.27\textwidth}
         \centering
                \includegraphics[width=0.95\columnwidth]{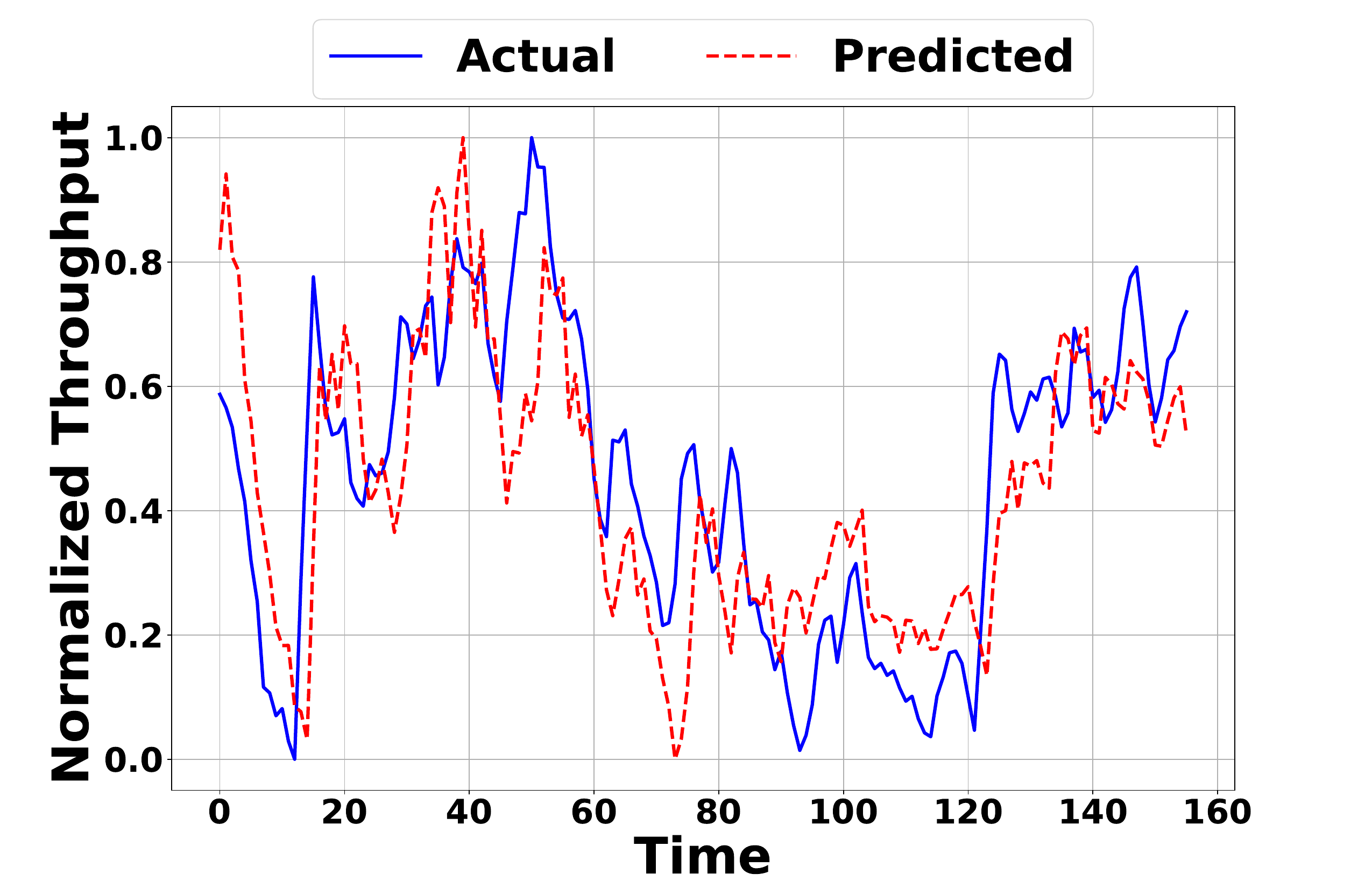}
	\caption{CNN+\fedbn{}}
	\label{fig:Trace_CFBN}
     \end{subfigure}
      \hspace*{-0.8cm}       
     \begin{subfigure}{0.27\textwidth}
         \centering
 \includegraphics[width=0.95\columnwidth]{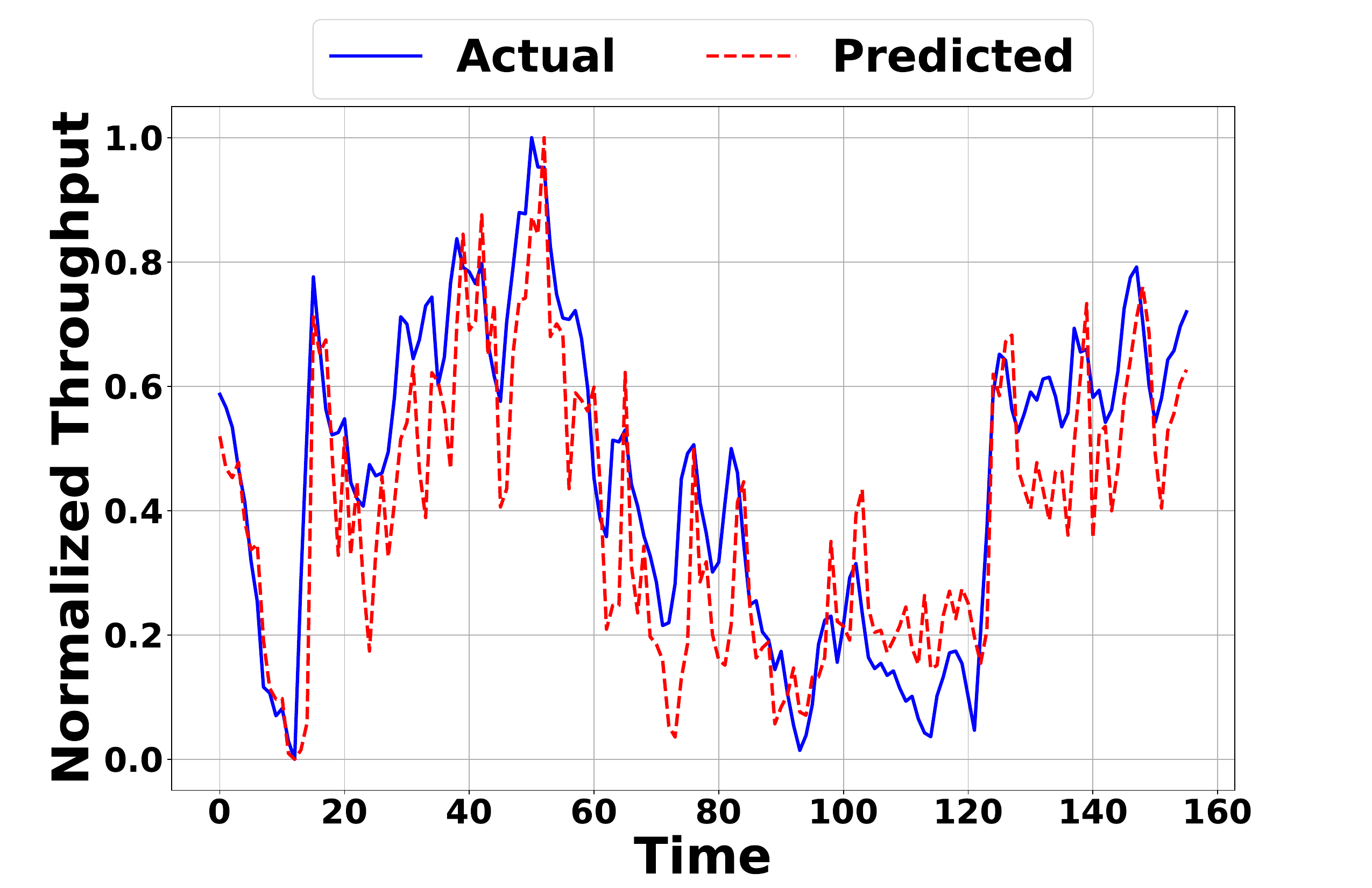}
	\caption{(\ac{LSTM}+\ac{CNN})+\fedbn{}}
	\label{fig:Trace_LCFBN}
     \end{subfigure}
     \hspace*{-0.8cm}       
     \begin{subfigure}{0.27\textwidth}
         \centering
\includegraphics[width=0.95\columnwidth]{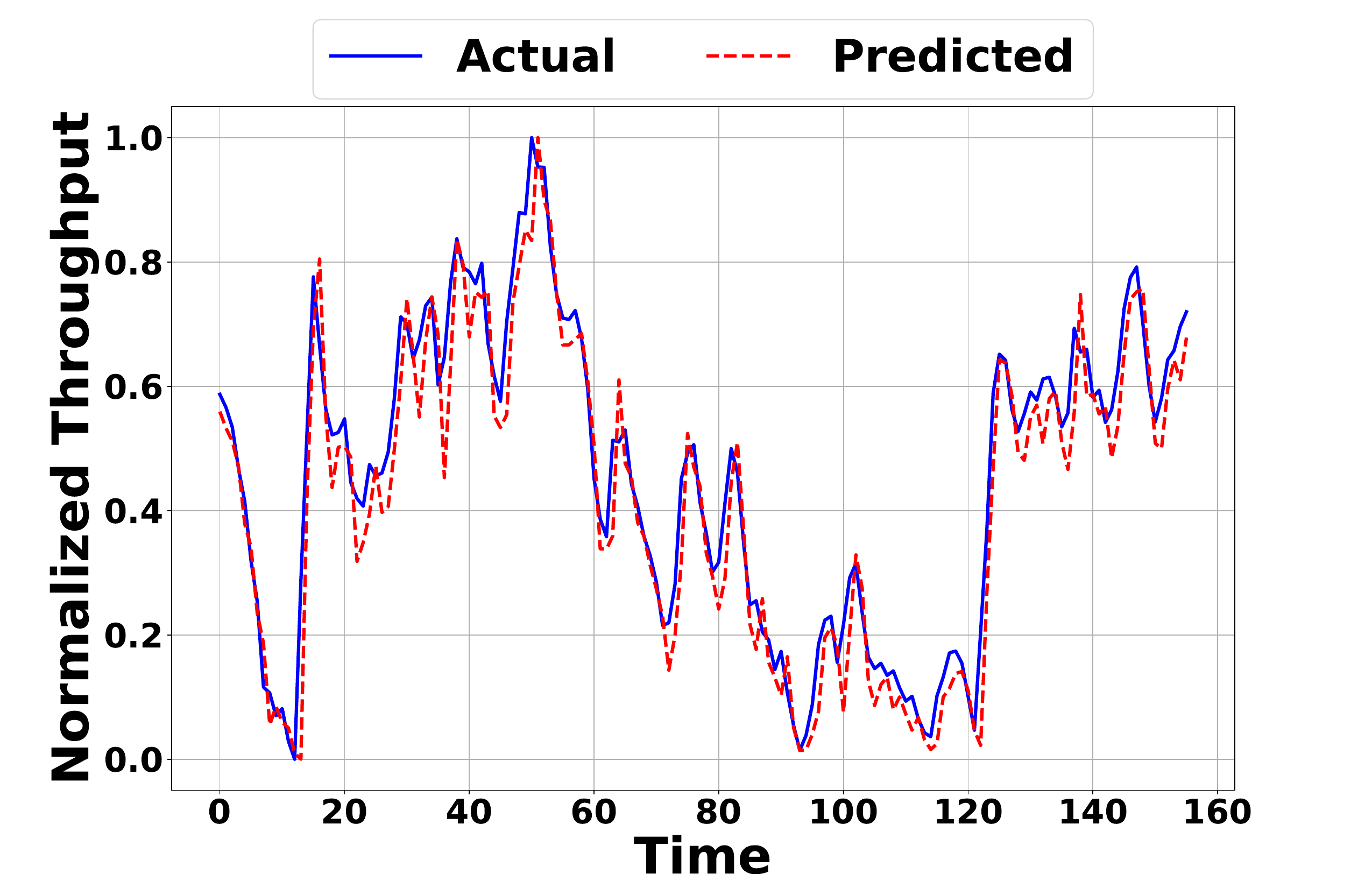}
	\caption{LSTM+\fedbn{}}
	\label{fig:Trace_LFBN}
     \end{subfigure}
     \hspace*{-0.8cm}       
     \begin{subfigure}{0.27\textwidth}
         \centering
\includegraphics[width=0.95\columnwidth]{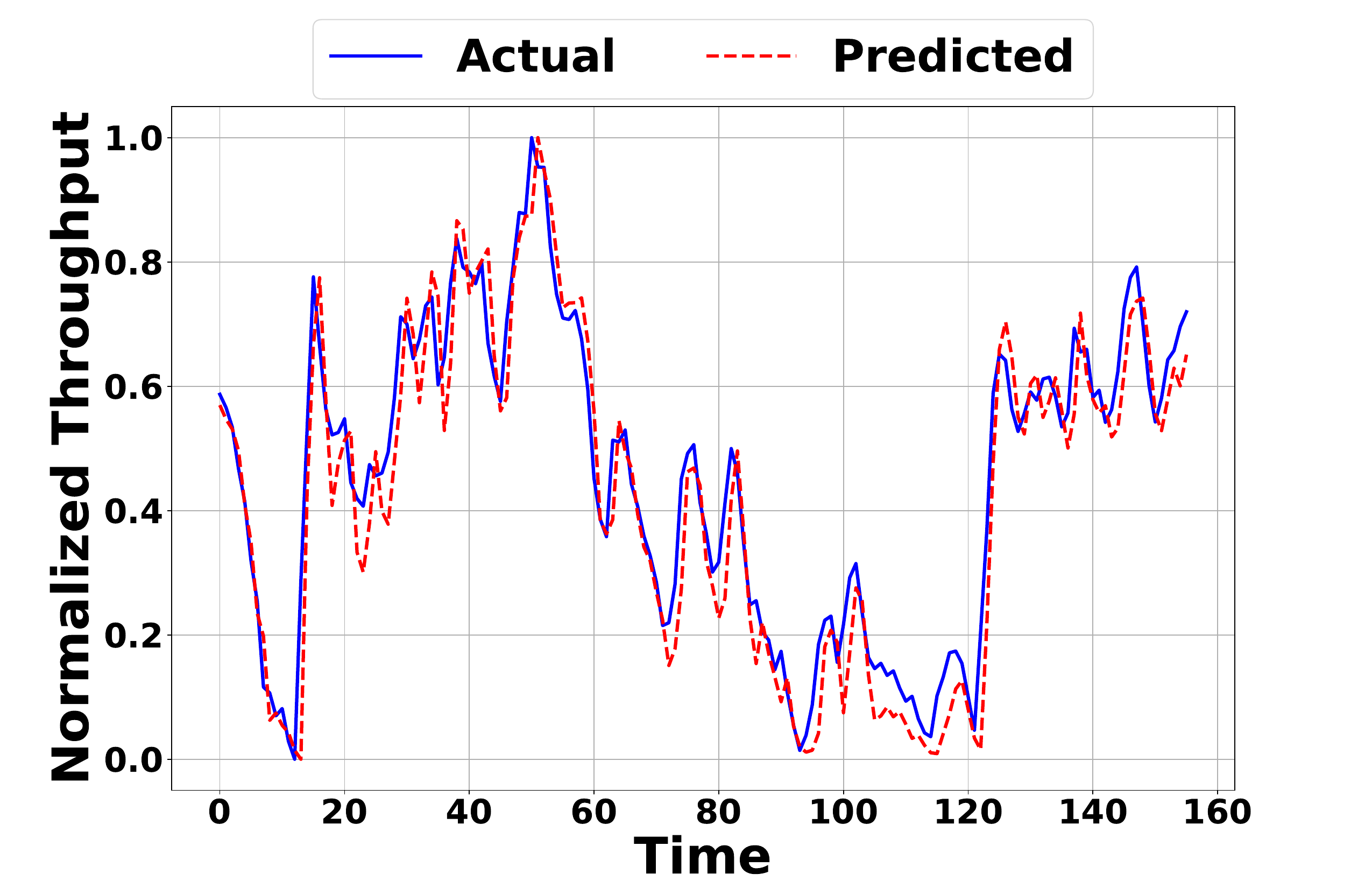}
	\caption{Transformer+\fedbn{}}
	\label{fig:Trace_TFBN}
     \end{subfigure}
       \caption{The Predicted (dotted lines) and the Ground (solid lines) throughput traces for Client 4 belonging to the dataset MNWild-TSA (Historical Window Size H=15 seconds, Prediction Window Size F=1 seconds). }
        \label{fig:thpttraces}
\end{figure*}

In this section, we present the implementation details of the proposed FL based 5G throughput prediction framework.
\subsection{Benchmarking with \textbf{Flower} and \textbf{Ray}}
\label{sec_methodology}
The overall benchmarking setup is implemented using the \textbf{Flower}\footnote{ \label{foot:flower} \url{https://flower.ai/docs/framework/index.html}}
FL framework (v1.8.0), which provides a flexible, framework-agnostic interface for experimenting with various model architectures and aggregation strategies. To enable parallel client execution, we integrate \textbf{Ray}\footnote{ \label{foot:ray} \url{https://www.ray.io}}  (v2.6.3), a distributed computing framework that schedules and executes Python tasks across multiple processes. All experiments are conducted on an Acer Nitro 5 (2021) with an AMD Ryzen 5 5600 CPU, 16GB RAM, and an NVIDIA RTX 3060 GPU (6GB).

\subsection{Hyperparameters and Generating Federated Clients}
In our setup, each federated 5G client corresponds to a throughput trace file from the publicly available datasets discussed in Section~\ref{sec:data_analysis}. These datasets, collected across diverse environments using different mobile devices, exhibit non-IID characteristics. Using \textbf{Flower}, each trace file is instantiated as a separate federated client, resulting in twenty clients in total: five each from MNWild-Ver, Lumos-5G, and Irish, four from MNWild-TSA, and one from MNWild-TNSA.

Each client uses an 80:20 split of its local data for training and testing, respectively. The client initializes its model during inference with the current local and global weights. The held-out test set \( D_{\text{test}} \), containing preprocessed historical network features and throughput values, is used to evaluate the predictive performance.
\subsection{Benchmarking across Model Architectures and Aggregation Strategies}

\newadd{
\tablename~\ref{tab:model-config} summarizes the model architectures and their respective training hyperparameters used in the experiments.}
We have executed the training over 100 global rounds, with local epoch counts being determined by each model architecture. In each round, 85\% of clients are randomly selected for training, and all clients are evaluated locally. Each client is assigned one CPU core and one GPU core to ensure isolated execution.

\section{Evaluation}\label{sec:results}
\begin{figure*}[h]
\centering 
     \begin{subfigure}{0.27\textwidth}
         \centering
                \includegraphics[width=0.95\columnwidth]{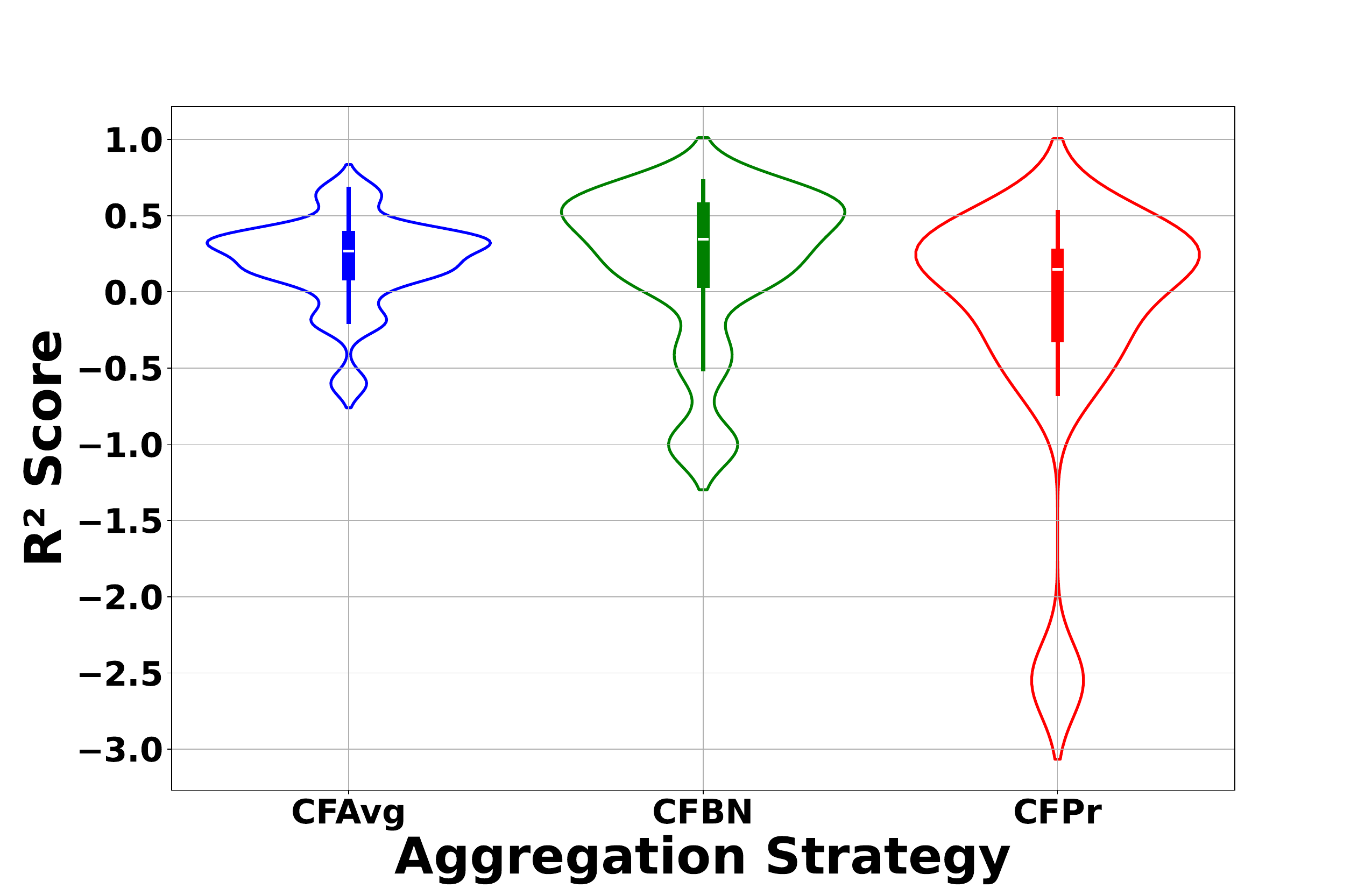}
	\caption{CNN}
	\label{fig:Choice_Aggregation_CNN}
     \end{subfigure}
      \hspace*{-0.8cm}       
     \begin{subfigure}{0.27\textwidth}
         \centering
 \includegraphics[width=0.95\columnwidth]{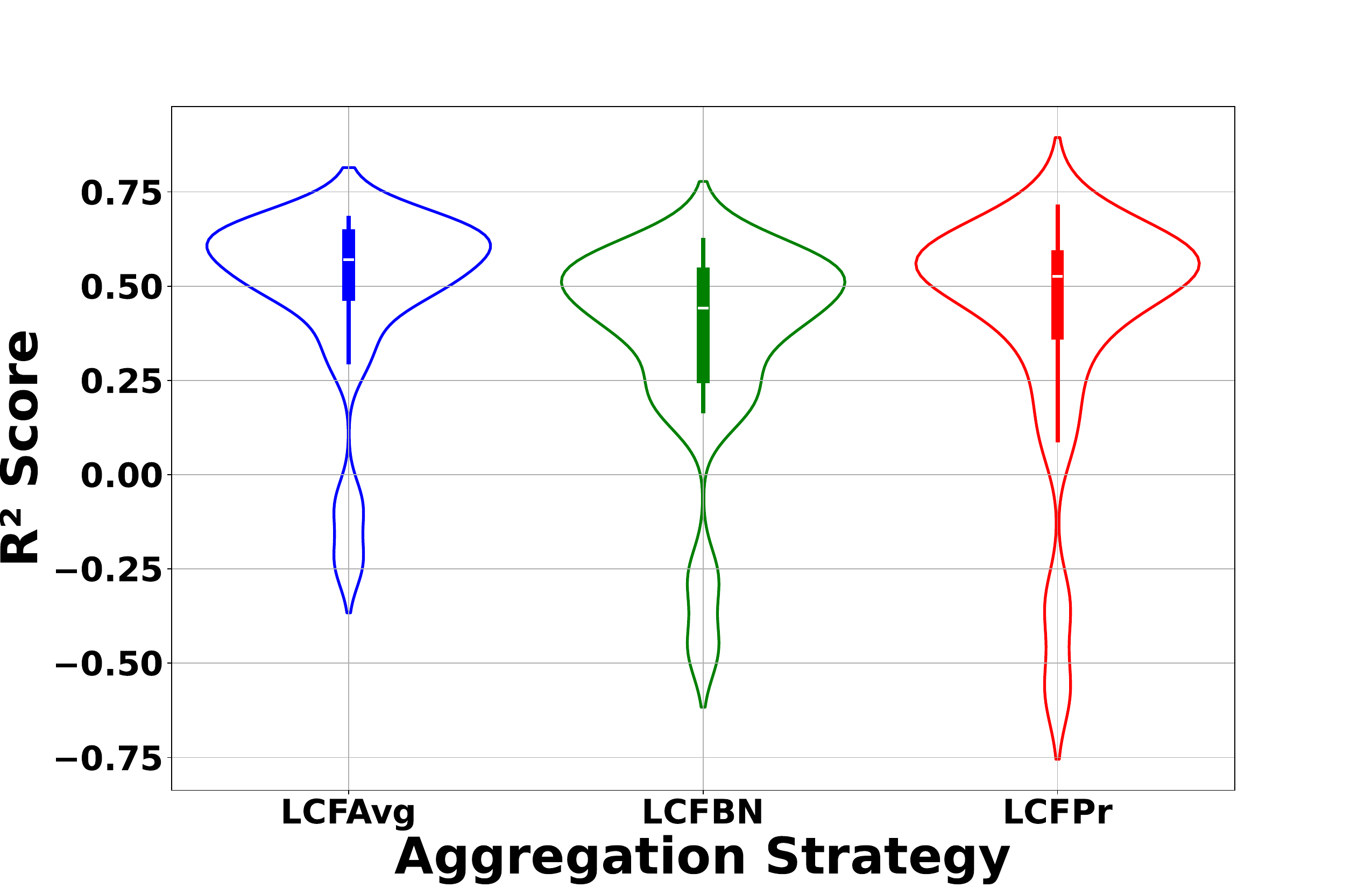}
	\caption{\ac{LSTM}+\ac{CNN}}
	\label{fig:Choice_Aggregation_LCNN}
     \end{subfigure}
     \hspace*{-0.8cm}       
     \begin{subfigure}{0.27\textwidth}
         \centering
\includegraphics[width=0.95\columnwidth]{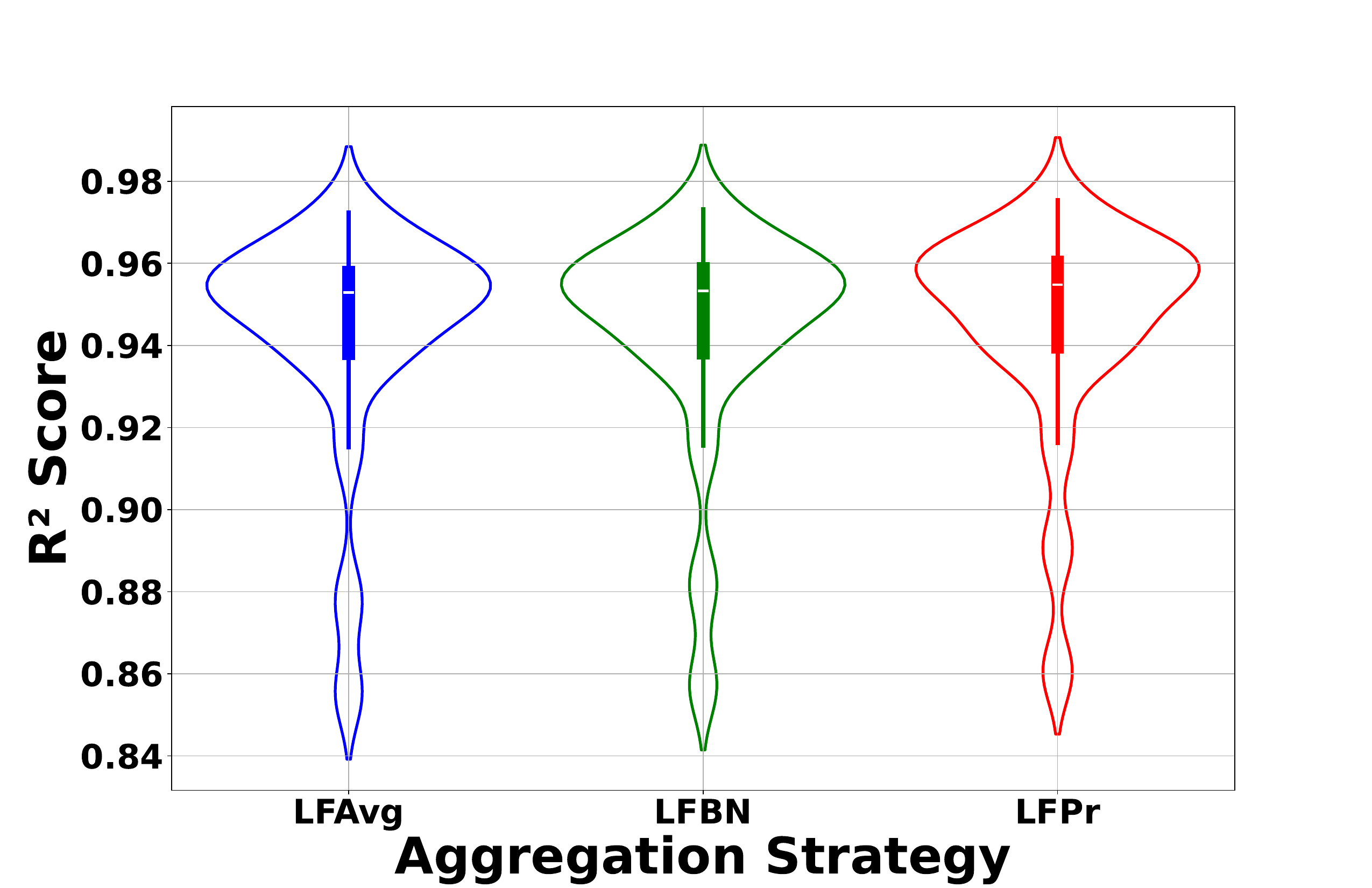}
	\caption{LSTM}
	\label{fig:Choice_Aggregation_LSTM}
     \end{subfigure}
     \hspace*{-0.8cm}       
     \begin{subfigure}{0.27\textwidth}
         \centering
 \includegraphics[width=0.95\columnwidth]{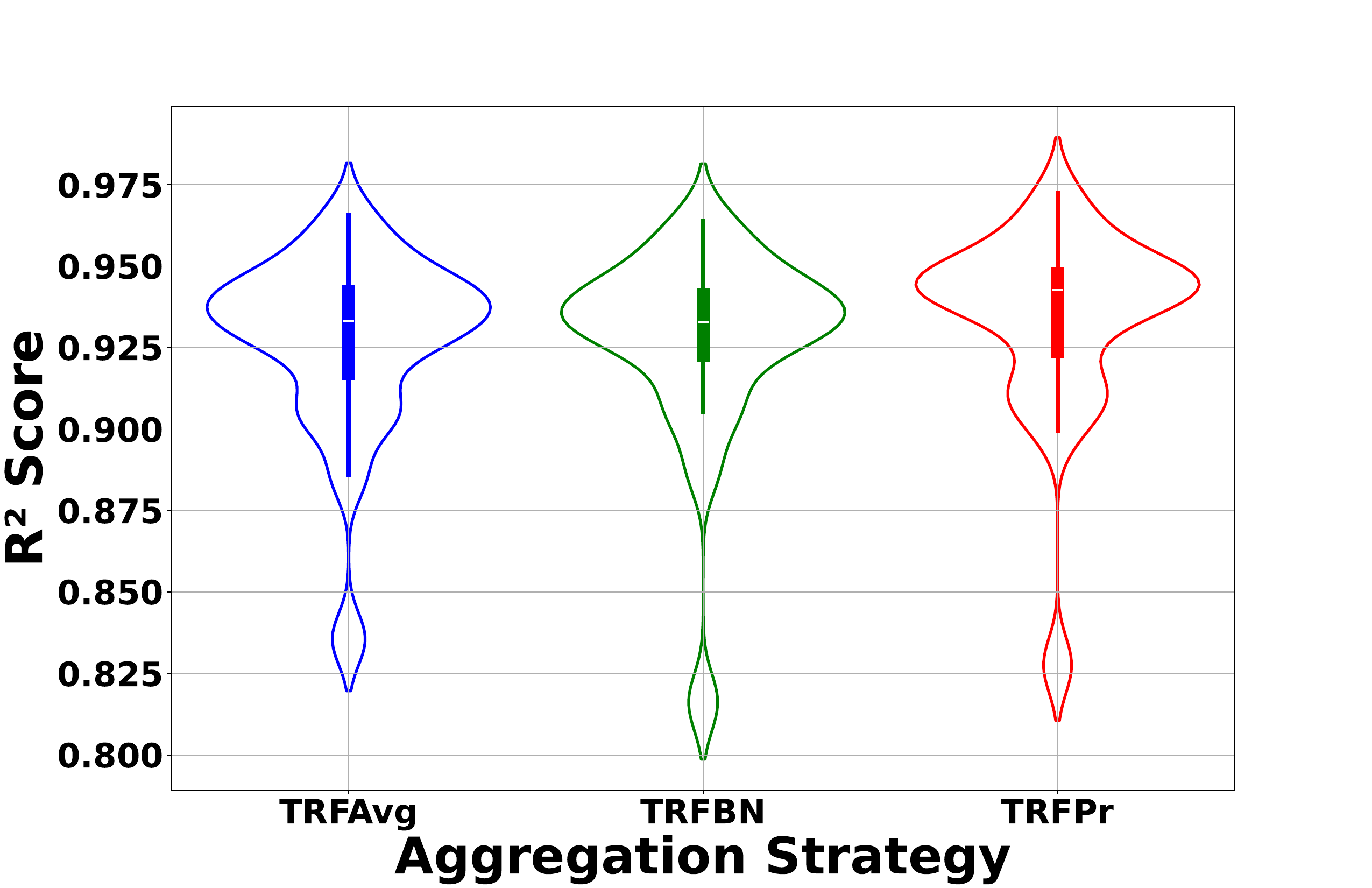}
	\caption{Transformer }
	\label{fig:Choice_Aggregation_TRN}
     \end{subfigure}
       \caption{Impact of the Federated Aggregation Strategies, with H=15 secs and F=1 sec.}
        \label{fig:Aggregation}
\end{figure*}
 In this work, in addition to investigating the federated throughput prediction framework, we have also developed a proof-of-concept to show how the framework can be used to choose optimal video bitrates of a live streaming application. In this section, we first present the performance analysis of the federated throughput prediction framework and subsequently discuss the efficacy of the federated throughput prediction in live-streaming applications.
 \subsection{Performance Analysis of Federated Throughput Prediction}
 \begin{figure}[t]
\centering 
     \begin{subfigure}{0.25\textwidth}
         \centering         \includegraphics[width=0.95\columnwidth,trim={1cm 0cm 4cm 6cm}]{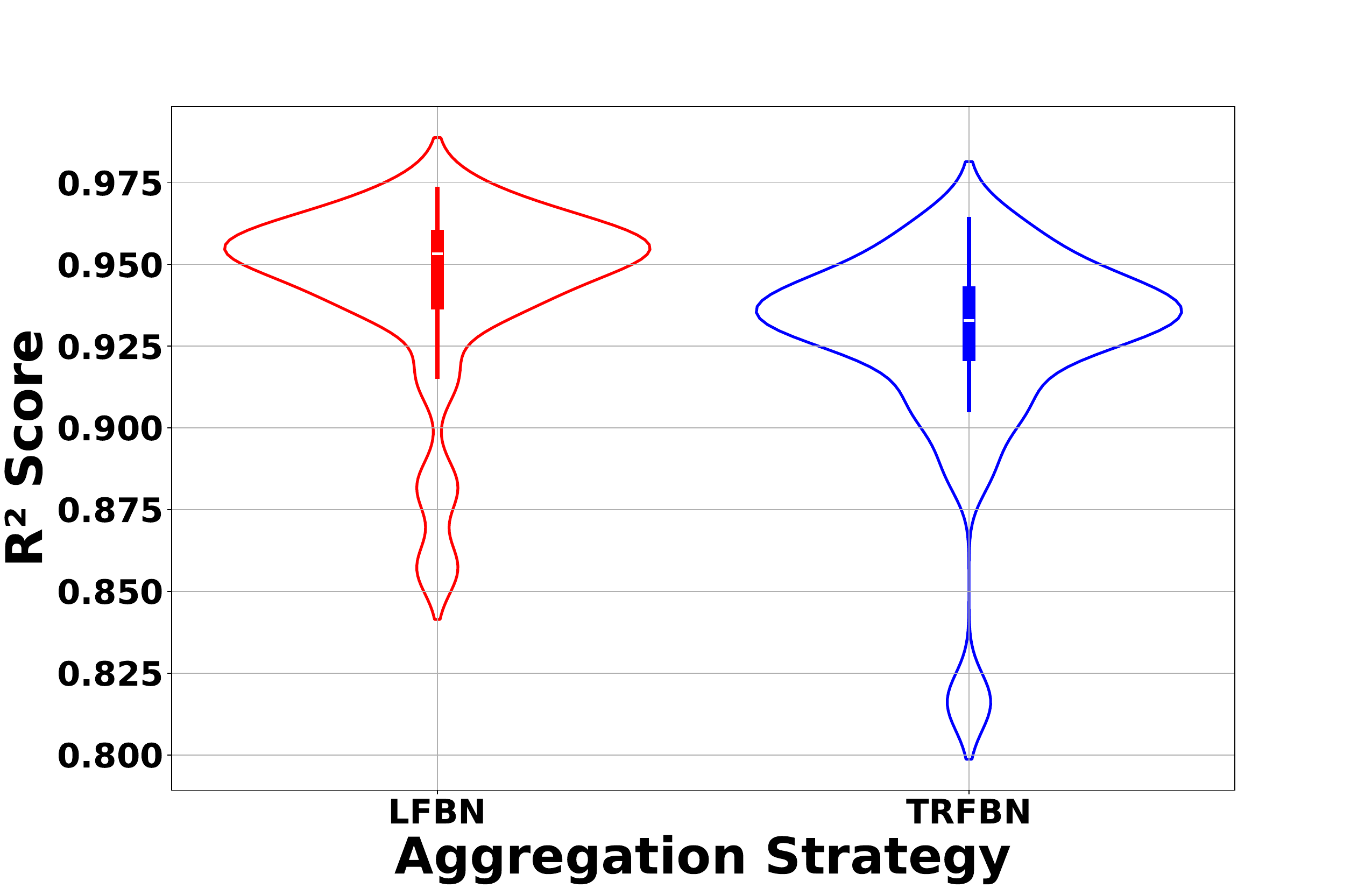}
	\caption{$R^2$ scores}
	\label{fig:r2_score}
     \end{subfigure}
      \hspace*{-0.5cm}       
     \begin{subfigure}{0.25\textwidth}
         \centering
\includegraphics[width=0.87\columnwidth,trim={1cm 0cm 4cm 6cm}]{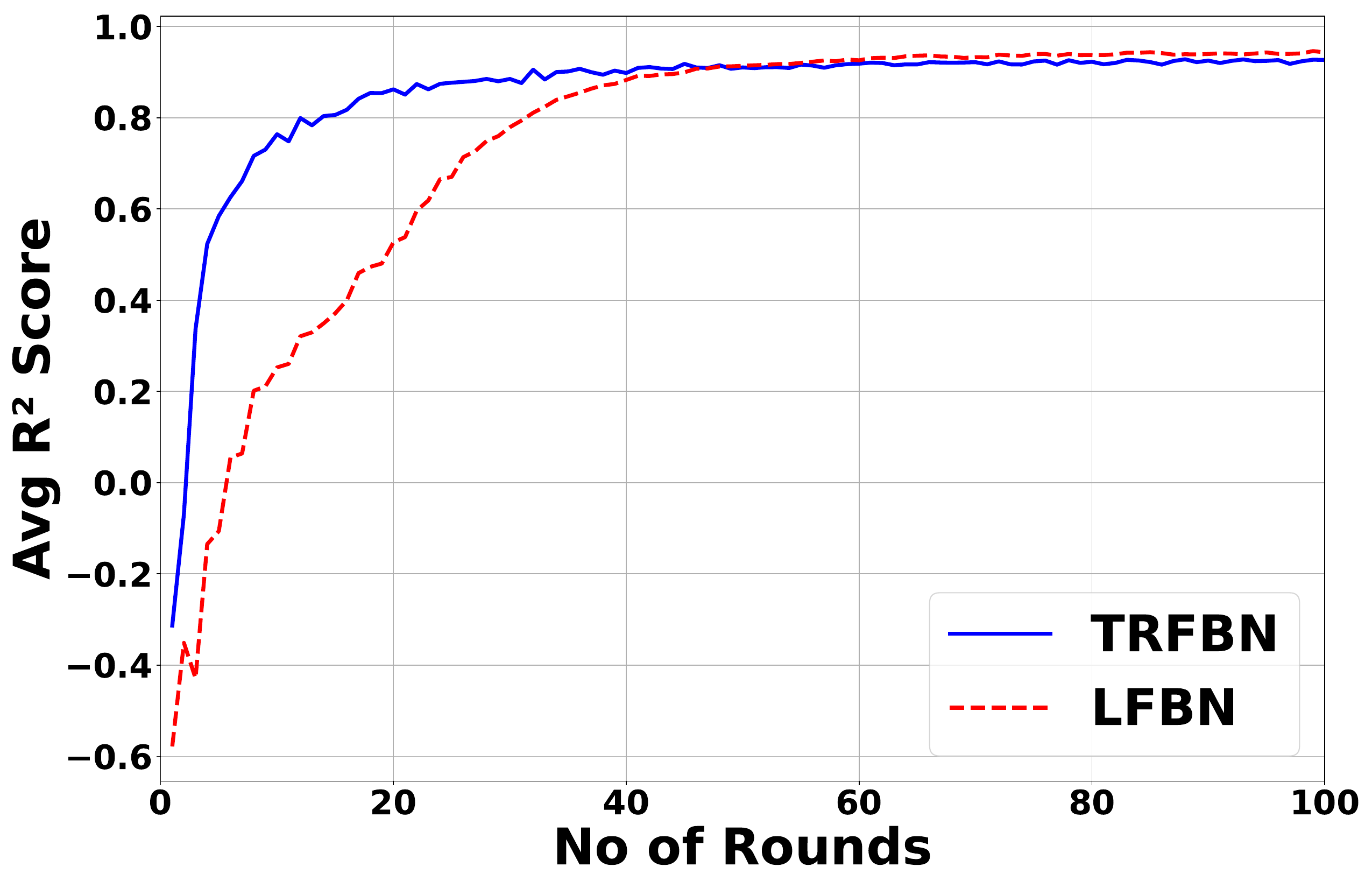}
	\caption{Convergence}
	\label{fig:convergence}
     \end{subfigure}
       \caption{Comparison of LSTM and Transformer models across global rounds with FedBN, (with $H=15$ secs, $F=1$ sec).}
        \label{fig:LSTMvsTrans}
\end{figure}

\subsubsection{\textbf{Choice of Model Architectures}}
\figurename~\ref{fig:thpttraces} shows the true throughput values as well as the predicted throughput traces corresponding to the four model architectures, \ac{CNN}, \ac{LSTM}+\ac{CNN}, \ac{LSTM}, and Transformers, trained with the \fedbn{} aggregation strategy of one federated cohort (Client 4 of the MNWild TSA dataset). We have considered a historical window of $H = 15$ seconds, a prediction horizon of $F = 1$ second, and a moving average filter window of $W = 3$ seconds. It may be observed that both the LSTM and the Transformer models closely predict the actual throughput with $R^2$ scores of 0.938 and 0.933, respectively. On the other hand, the mean $R^2$ scores of CNN and the hybrid \ac{LSTM}+\ac{CNN} models across the twenty cohorts are 0.51 and 0.622, respectively. {The reason may be attributed to the internal memory of LSTM and the self-attention mechanism of Transformers, which enable them to remember the time dependencies. On the other hand, CNN possinly fails to capture the temporal dependency as it only learn patterns and not the time-varying relations.} So, we continue the rest of our discussion using LSTM and Transformers.

\newadd{A comparison between LSTM and Transformer models in the federated setting, as illustrated in \figurename~\ref{fig:r2_score} and \figurename~\ref{fig:convergence}, shows that while both architectures achieve similar peak $R^2$ scores (0.938 for LSTM and 0.933 for Transformer), the convergence behavior differs significantly. Specifically, the Transformer model reaches its performance peak within 10 global rounds, whereas the LSTM model requires nearly 40 rounds to achieve the same level of accuracy. However, the Transformer also exhibits higher variance across training rounds. These observations suggest that Transformer models may be more suitable for time-sensitive, real-time applications due to their rapid convergence. In contrast, LSTM models are preferable when stability and reliability are prioritized.}


\subsubsection{\textbf{Choice of Aggregation Strategy}}
\newadd{\figurename~\ref{fig:Aggregation} presents violin plots of the $R^2$ scores for the twelve combinations of model architectures (CNN, \ac{LSTM}+\ac{CNN}, LSTM, Transformer) and aggregation strategies (\fedavg{}, \fedbn{}, \fedprox{}) with historical window \(H=15\) seconds and forecast window \(F=1\) second. The results reveal several key trends. For the CNN architecture (\figurename~\ref{fig:Aggregation}a), the use of \fedbn{} results in a visibly higher mean $R^2$ score and a significantly narrower interquartile range compared to both \fedavg{} and \fedprox. This suggests that \fedbn{} is more effective in mitigating variance induced by non-IID client distributions in shallow convolutional models. In contrast, for the \ac{LSTM}+\ac{CNN} hybrid (\figurename~\ref{fig:Aggregation}b), the highest mean $R^2$ score is achieved under \fedavg{}, although \fedbn{} again demonstrates a tighter distribution, indicating improved stability across clients. With deeper temporal models (\figurename~\ref{fig:Aggregation}c–d), like LSTM and Transformer, the differences between aggregation strategies become marginal. All three strategies yield mean $R^2$ scores within a narrow band: approximately 0.938 for LSTM and 0.933 for Transformer. While the central tendency is comparable across strategies, \fedbn{} exhibits slightly lower variance, further reinforcing its robustness under heterogeneity, mainly when used with more expressive models. These observations indicate that the aggregation strategy plays a more prominent role for shallower architectures (e.g., CNN), whereas deeper temporal models (LSTM, Transformer) are inherently more robust to heterogeneity. Consequently, in the remainder of our experiments, we fix the aggregation strategy to \fedbn{} to leverage its consistently lower variance across model types.}

\subsubsection{\textbf{Parametric Analysis of History and Future Window}}
\begin{figure}[t]
\centering 
     \begin{subfigure}{0.25\textwidth}
         \centering
         \includegraphics[width=0.96\columnwidth,keepaspectratio]{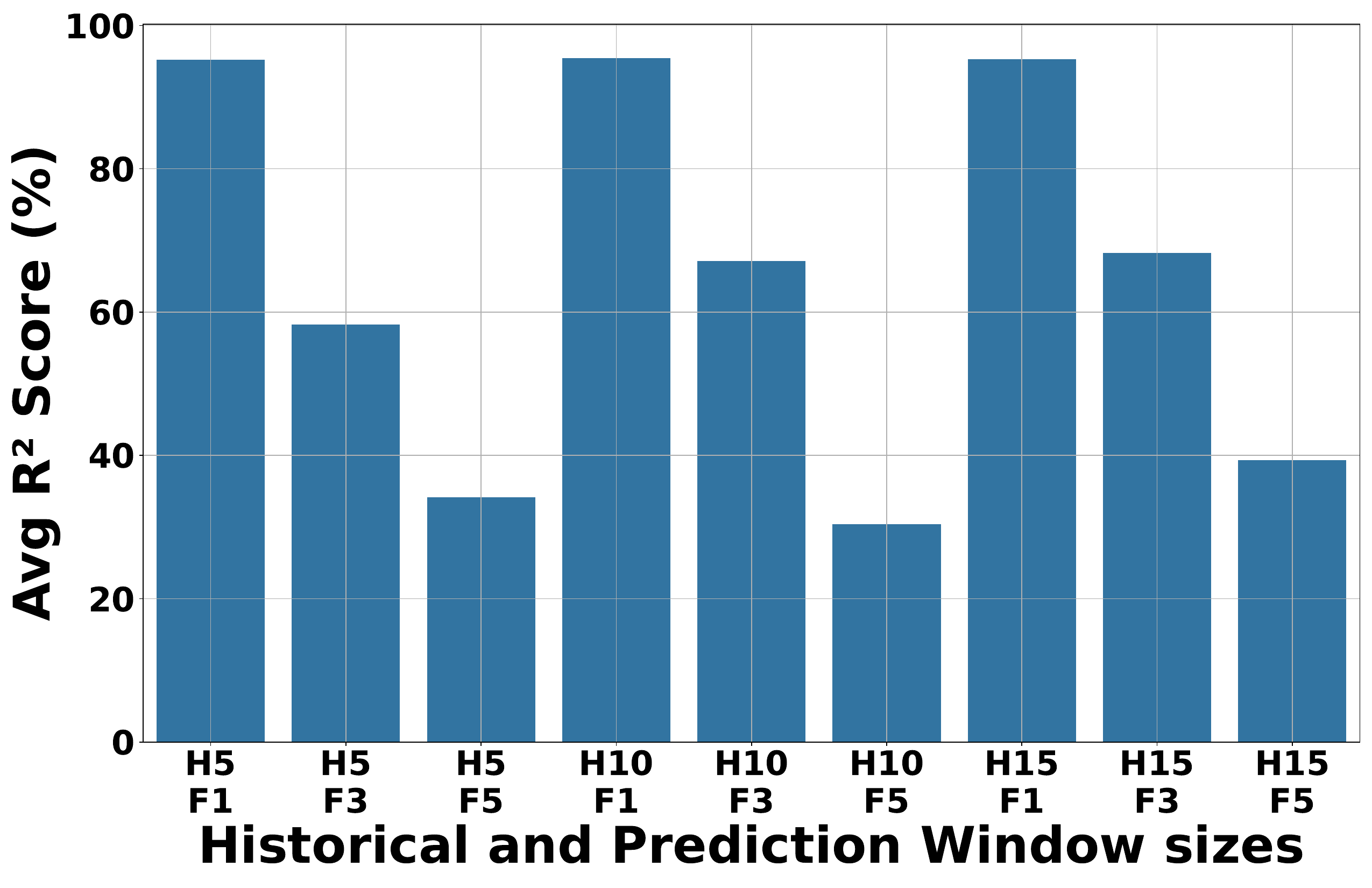}
	\caption{}
	\label{fig:Window size R2 score_L}
     \end{subfigure}
      \hspace*{-0.5cm}       
     \begin{subfigure}{0.25\textwidth}
         \centering
\includegraphics[width=0.95\columnwidth,keepaspectratio]{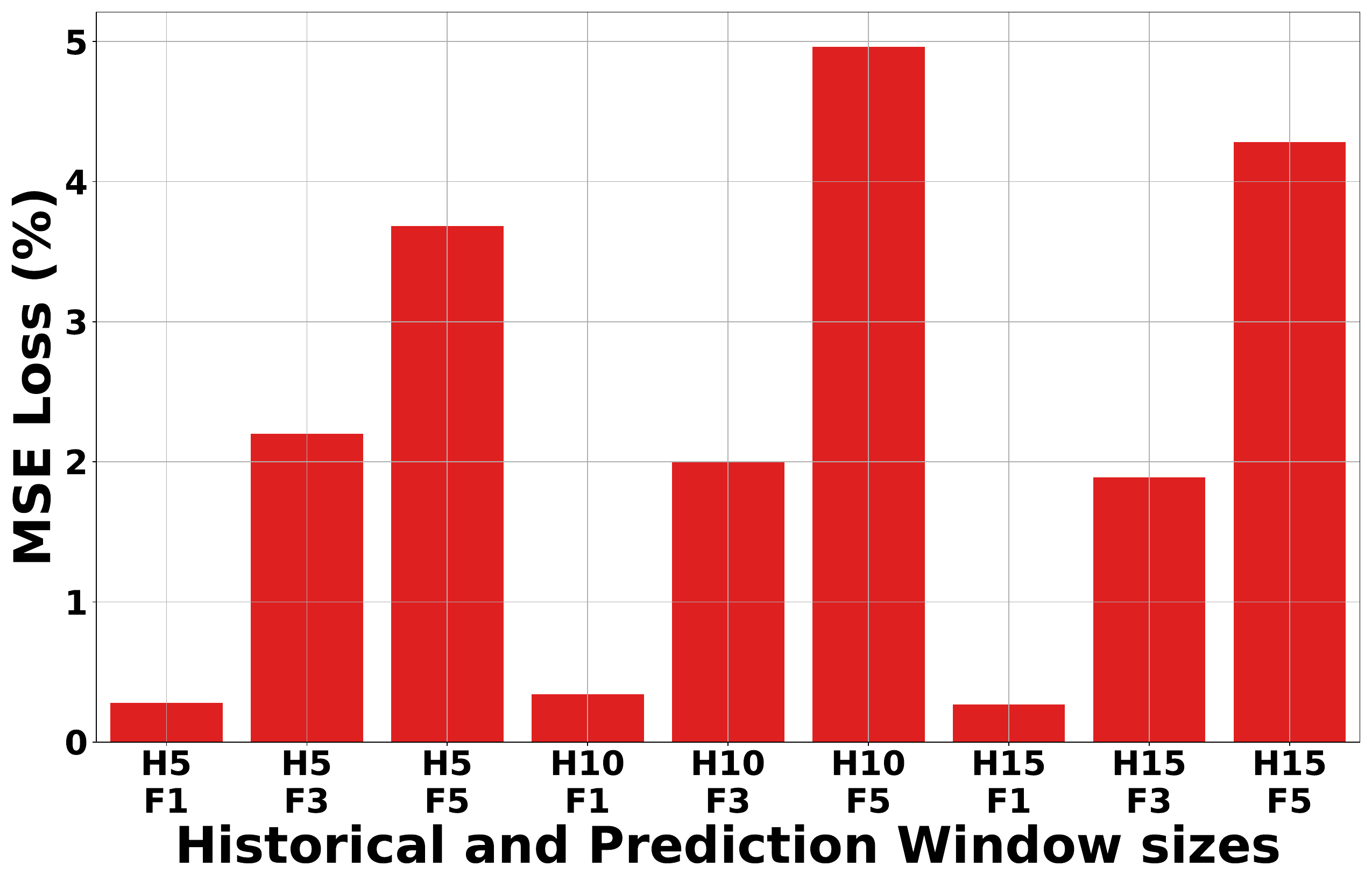}
	\caption{}
	\label{fig:MSE_L}
     \end{subfigure}
       \caption{Analysis of LSTM+\fedbn{} for different history and future window size: (a) $R^2$-Score (b) MSE Loss.}
        \label{fig:LSTM_FedBN}
    \end{figure}
\begin{figure}[t]
\centering 
     \begin{subfigure}{0.25\textwidth}
         \centering
         \includegraphics[width=0.95\columnwidth,keepaspectratio]{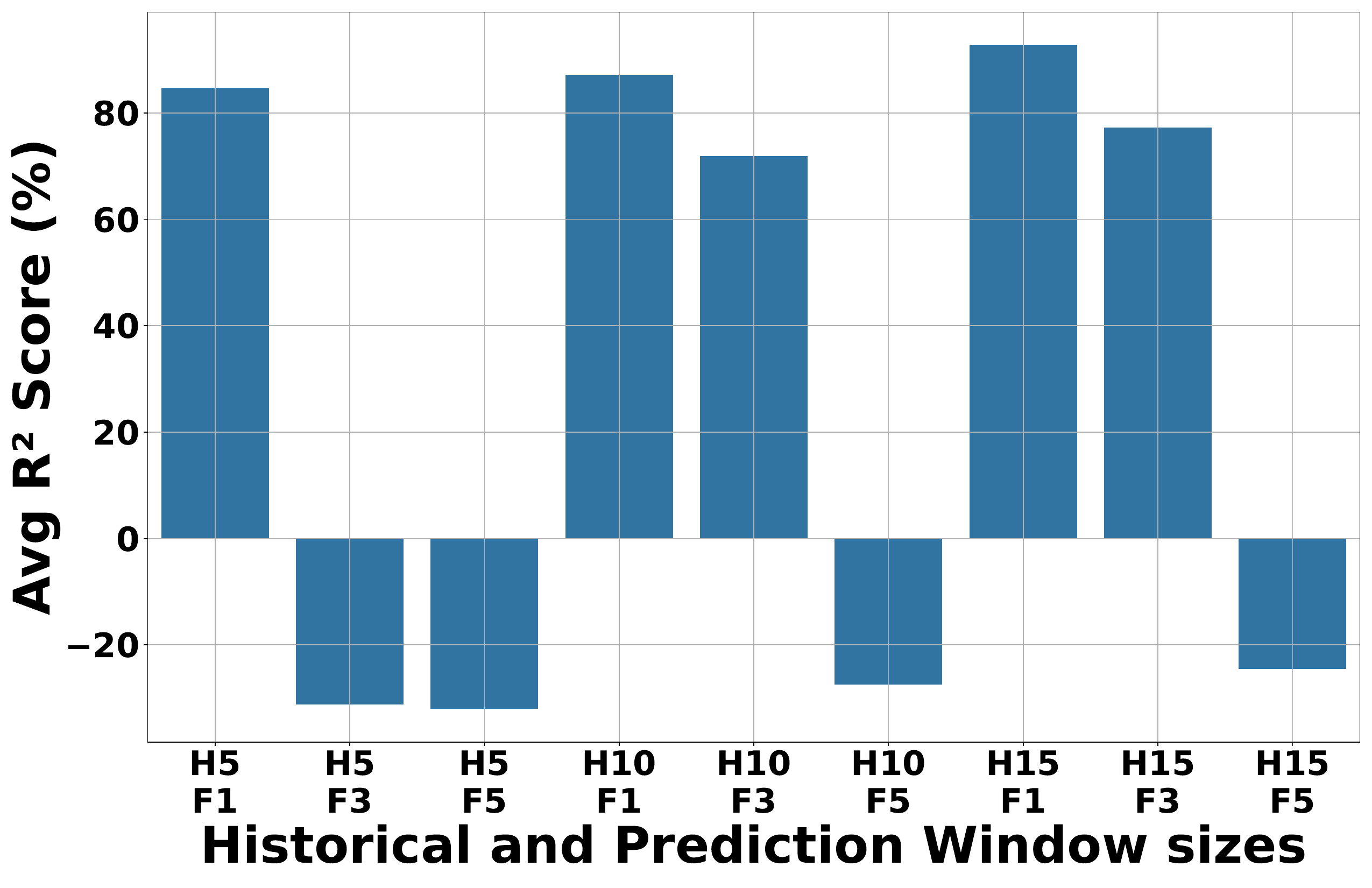}
	\caption{}
	\label{fig:Window size R2 score_T}
     \end{subfigure}
      \hspace*{-0.5cm}       
     \begin{subfigure}{0.25\textwidth}
         \centering
\includegraphics[width=0.95\columnwidth,keepaspectratio]{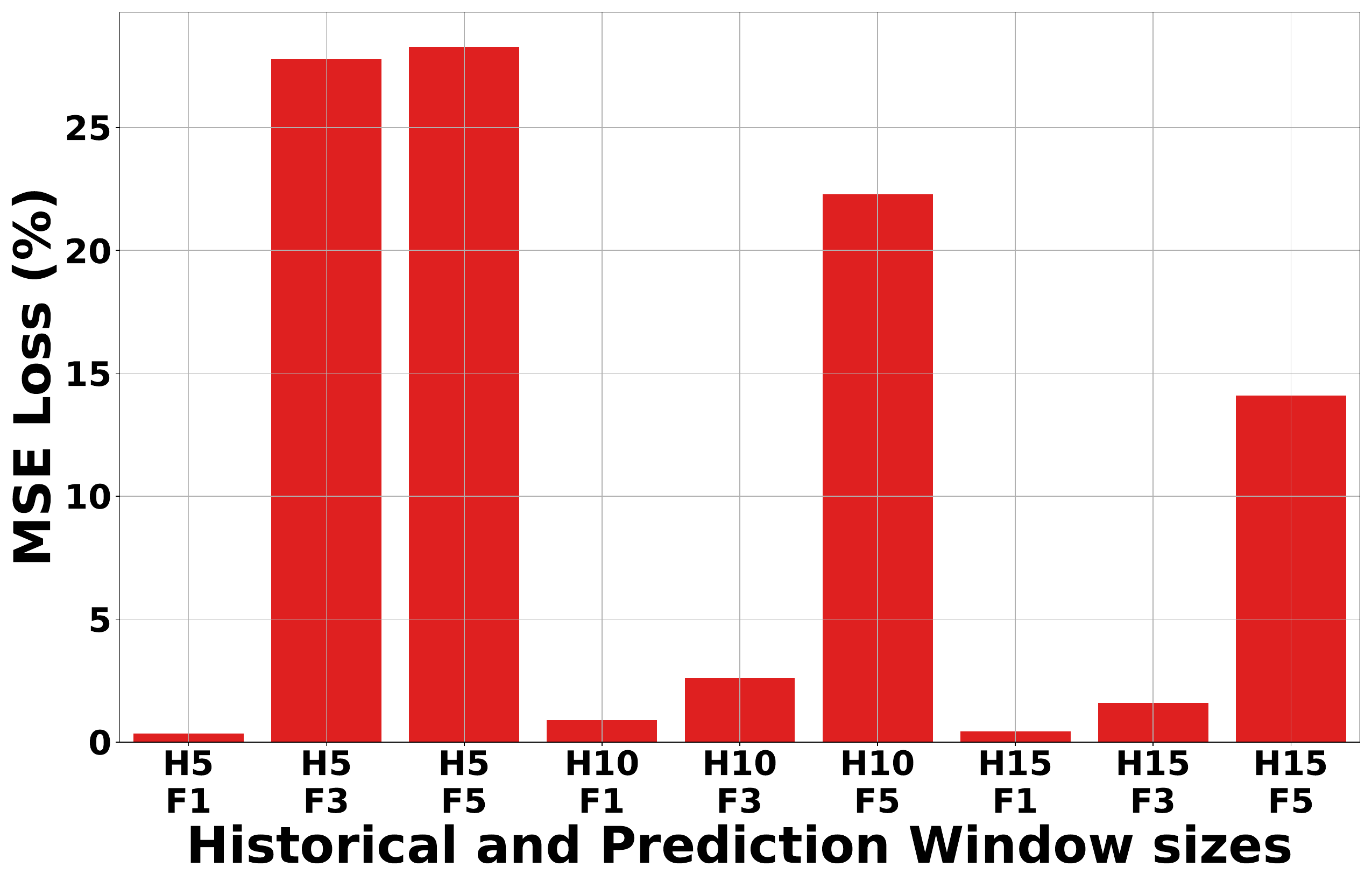}
	\caption{}
	\label{fig:MSE_T}
     \end{subfigure}
       \caption{Analysis of Transformer+\fedbn{} for different history and future window size: (a) $R^2$-Score (b) MSE Loss.}
        \label{fig:TransFedbn}
    \end{figure}
We next present a parametric analysis of the LSTM and Transformer models under the \fedbn{} aggregation strategy, varying both the history (H) and prediction (F) window sizes. \figurename~\ref{fig:LSTM_FedBN} and \figurename~\ref{fig:TransFedbn} show the corresponding average $R^2$ scores and MSE losses for the LSTM and Transformer models, respectively. Two observable trends are:
(i) For a fixed history window, the $R^2$ score tends to decrease, and \ac{MSE} increases as the prediction window grows.
(ii) For a fixed prediction window, increasing the history window generally improves the $R^2$ score and reduces \ac{MSE}. 
The first trend can be attributed to the reduced temporal correlation between the present network features and the future throughput over longer prediction horizons, leading to lower predictive accuracy. This is in agreement with~\cite{biswas2023kalman}. The second trend may be attributed to the fact that a longer historical window provides more data for training and inference, resulting in higher prediction accuracy.

An important observation is that Transformers require a longer historical window to achieve an acceptably high $R^2$ score and low \ac{MSE}. For instance, to exceed an $R^2$ score of 90\%, Transformers need $H=15$ seconds, whereas LSTM achieves this with $H=5$ seconds. This shows LSTM’s efficiency in capturing short-term temporal patterns.

For the rest of the paper, we show the results using LSTM and \fedbn{} only. 
\subsubsection{\textbf{Impact of Cohort Size}}
\begin{figure}[t]%
	\centering
        \includegraphics[width=0.7\columnwidth,trim={1cm 0cm 4cm 3cm}]{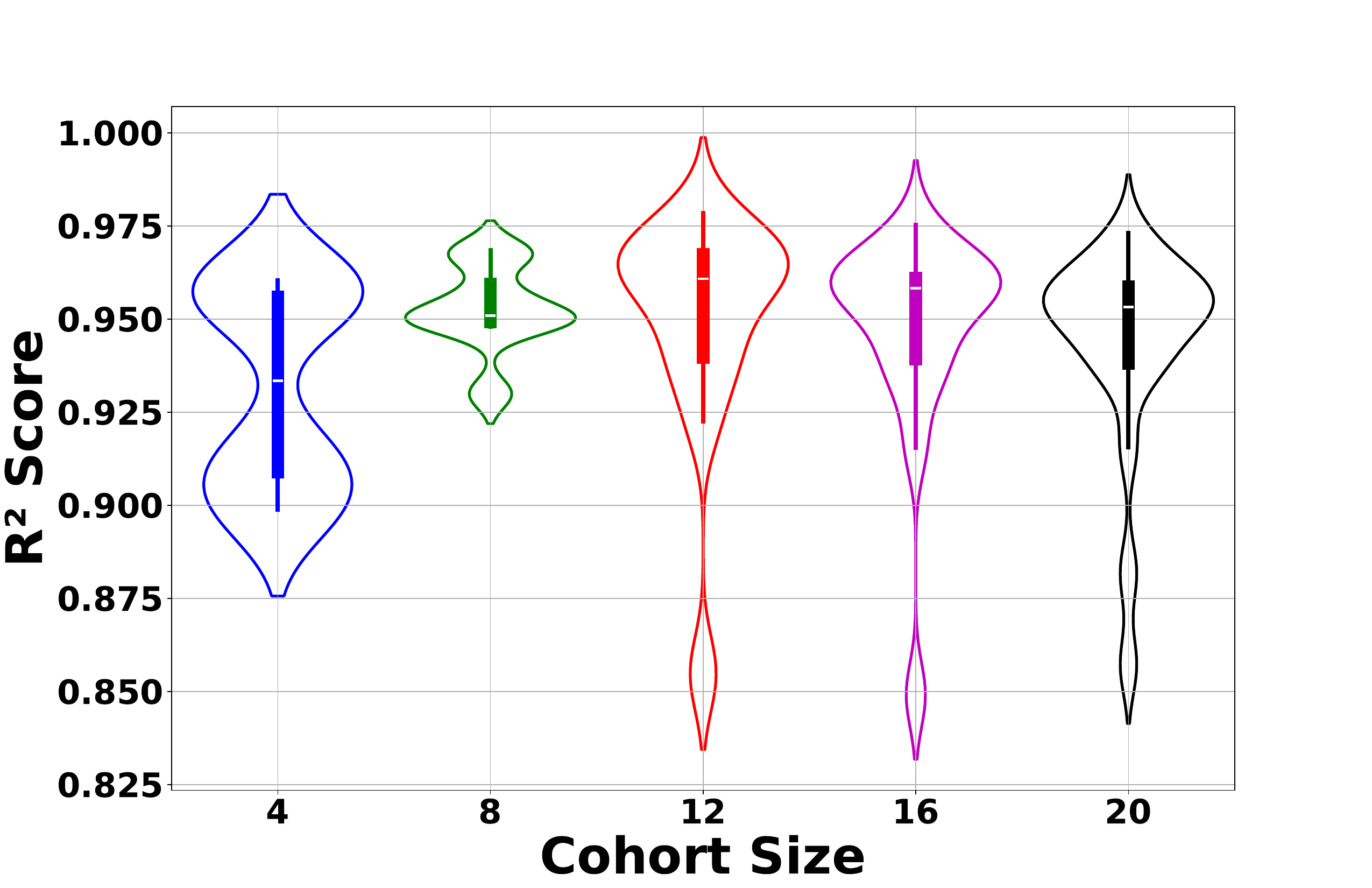}
	\caption{Distribution of $R^2$-Score across different cohort size of LSTM-\fedbn{}, taking one client from each dataset, two clients from each dataset, and so on, (H = 15 secs, F = 1 sec).}
	\label{fig:Cohort_Size_combination_1}
\end{figure}

We next examine the impact of cohort size on the performance of the LSTM+\fedbn{} algorithm. For this, we combine four clients from MNWild-TSA and one from MNWild-TNSA into a single group, denoted as MNWild-Tmobile. The participating clients are thus drawn from four datasets: MNWild-Ver, MNWild-Tmobile, Lumos-5G, and Irish, with five clients per dataset. \figurename~\ref{fig:Cohort_Size_combination_1} shows violin plots of $R^2$ scores for cohort sizes of 4, 8, 12, 16, and 20, where cohort size refers to the number of clients in training. The median and mean $R^2$ scores remain consistently high, confirming the robustness of LSTM+\fedbn{}. With 4 clients, the mean $R^2$ is 0.9315; increasing to 8 improves the mean to 0.9527 and reduces variance to 1.28$\times 10^{-4}$, indicating greater model stability. For 12, 16, and 20 clients, mean $R^2$ stays above 0.94, though variance slightly rises (still in the order of $10^{-3}$), likely due to increased heterogeneity. Thus, a moderate cohort size (e.g., 8) offers the best trade-off between accuracy and stability, yielding the most compact and reliable $R^2$ distribution.

\begin{figure}[t]%
	\centering
        \includegraphics[width=0.7\columnwidth,trim={1cm 0cm 4cm 3cm}]{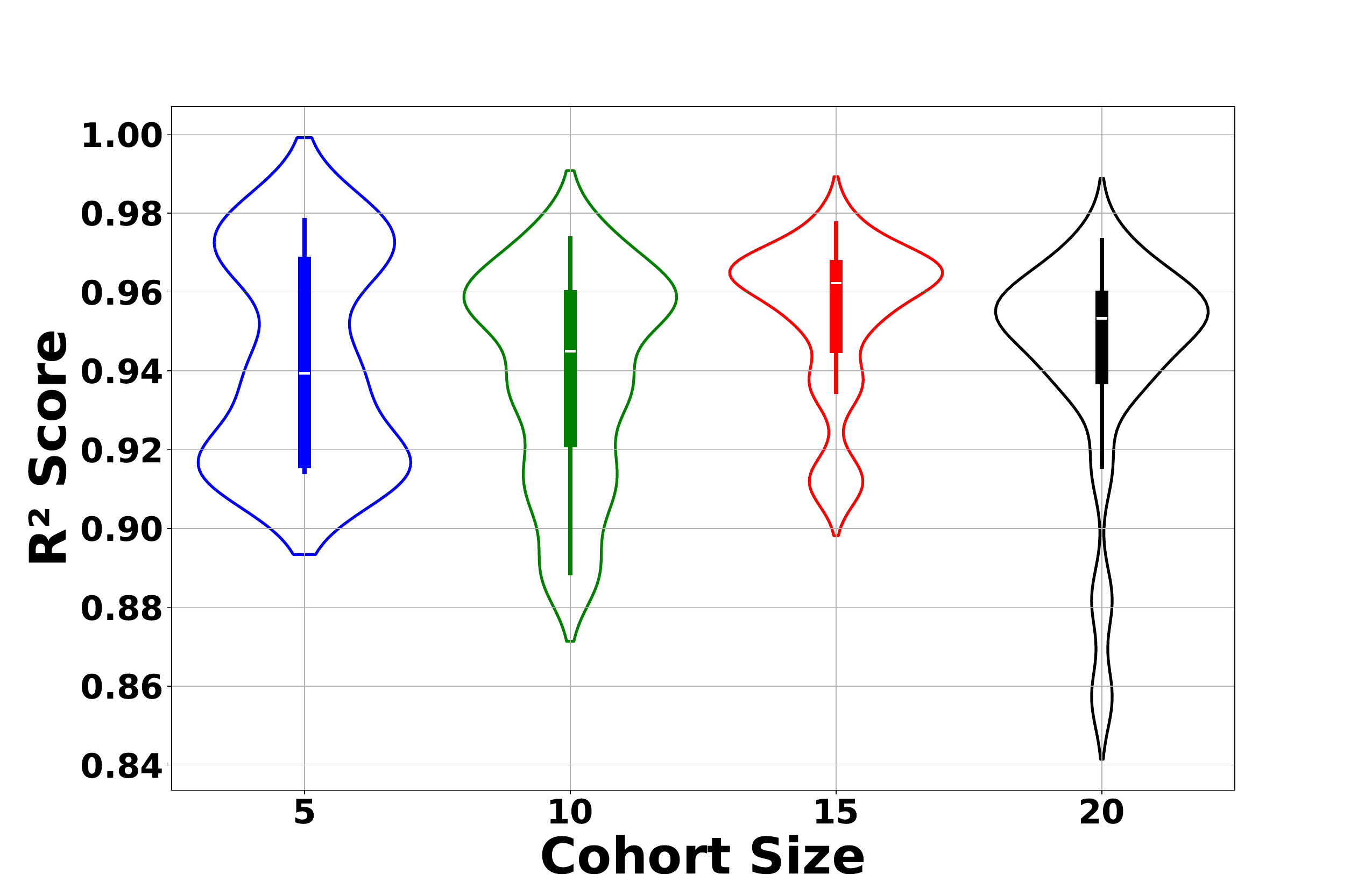}
	\caption{Distribution of $R^2$-Score across different cohort size of LSTM+\fedbn{} ( MNWild-Tmobile = 5 clients,  MNWild-Tmobile + MNWild-Ver = 10 clients,  MNWild-Tmobile + MNWild-Ver + Lumos-5G =15 clients,  MNWild-Tmobile + MNWild-Ver + Lumos-5G + Irish =20 clients).}
	\label{fig:Cohort_Size_combination_2}
\end{figure}

\newadd{\figurename~\ref{fig:Cohort_Size_combination_2} investigates the impact of a different combination of cohorts. The first cohort consists solely of clients from the MNWild-Tmobile dataset, followed by a combination with MNWild-Ver, and then sequential additions of Lumos-5G and Irish. The primary observation is that the median $R^2$ score remains consistently high across all combinations, suggesting that the model maintains robust predictive performance under varying cohort compositions. Quantitatively, the cohort of MNWild-Tmobile alone yields a mean $R^2$ score of 0.9432. The addition of MNWild-Ver slightly reduces the mean to 0.9391, likely due to complementary yet heterogeneous traffic patterns in the two subregions. Upon incorporating the Lumos-5G dataset, the mean $R^2$ improves notably to 0.9534. However, including the Irish dataset (20 clients) introduces higher geographical and infrastructural diversity, slightly reducing the mean to 0.9431. Overall, the results suggest that the LSTM+\fedbn{} model maintains stable predictive performance across different cohort compositions, with only marginal variations in mean $R^2$ and variance, indicating robustness to moderate levels of cross-regional heterogeneity.
}
\subsection{Proof-of-Concept: Live-Streaming Application}
In this section, we demonstrate how the FL-based throughput prediction can be used for live-streaming applications. 

In recent times, live-streaming has emerged as one of the most popular modes of video streaming. Live-streaming videos require ultra-reliable real-time guarantees, as they are recorded, segmented, encoded, and streamed in real-time. Therefore, a live-streaming video delivery over the mobile internet is highly susceptible to service disruptions due to the fluctuating network conditions. As mentioned in Section~\ref{sec:intro}, live-streaming uses the DASH protocol, where the {QoE} of live video users depends on the bitrates, bitrate changes, freezing events, and latency. The \ac{QoE} is maximized by an optimal selection of bitrates, without which the entire video delivery may get affected. For example, choosing a high bitrate in a network blindspot can cause a delayed delivery of the video segments, leading to a violation of the playback deadlines. As a result, the events of \textit{video stall}, and  the \textit{end-to-end playback latency} can increase. So, a live-streaming session at a specific edge device can benefit significantly from accurate throughput prediction, especially if the prediction caters to personalization.   

The \ac{QoE} of live-streaming videos can be written as~\cite{LSQoESun2019}:

 \begin{align}\label{eq:qoelive}
     QoE_\mathrm{live}  = & \sum_{i=1}^{\mathcal{N}_v}\mu_1 Q(R_i) - \mu_2 t^\mathrm{stall}_{i} \nonumber\\ 
     &-\mu_3\left|Q(R_{i+1})-Q(R_{i}) \right|-\mu_4\psi(l_i)- \mu_5\eta_i,
 \end{align}
Here, $R_i$ in bps is the video rate of segment $i$, and $Q(r)=log(r/R_{min})$ gives the perceptible video quality. The stall time is represented by $t^\mathrm{stall}_{i}$, which is a function of the total download time in seconds and the client-side buffer length in seconds. The download time in turn is a function of: (1) the video rate, (2) the client-side throughput in bits per second, (3) the segment duration, and (4) the overhead associated with the round-trip delay. The playback latency given by $\psi(l_i)$ is a logistic growth function  $\frac{1}{1+e^{\omega-l_i}}-\frac{1}{1+e^{\omega}}$, where $\omega$ gives the latency sensitivity range.  $\mu_1, \ \mu_2, \ \mu_3,\ \mu_4, \ \mu_5$ in (\ref{eq:qoelive}) denotes the sensitivity of the users to the different \ac{QoE} components. These values may be tuned to maximize the \ac{QoE}. The live-streaming algorithms in~\cite{LSQoESun2019} adopt a few assumptions. For example, it downloads the segments sequentially while ensuring that a segment is downloaded only after it is completely encoded. 

In this work, we have used the \textit{MPC-Chunk} algorithm of ~\cite{LSQoESun2019} to establish the use of FL-based throughput prediction for live-streaming. It breaks a video segment into chunks of shorter time lengths, such that the chunks can be sent directly to the clients as and when they are encoded. The bitrate of the chunks is chosen using model predictive control, which in turn requires the predicted network throughput and the round-trip time. In~\cite{LSQoESun2019}, the throughput has been predicted using harmonic mean, which we have replaced in this paper with the federated throughput prediction. 

\subsubsection{\textbf{Simulation Parameters}}
In this work, the live-streaming video application runs at each edge client, where throughput prediction takes place according to the proposed framework. We have used the live-video streaming simulator\footnote{\url{https://github.com/monkeysun555/benchmark\_mpc} (Last access: \today)} of~\cite{LSQoESun2019}. As it was designed to work with 4G network throughput, we have modified it to work with predicted 5G throughput traces by upscaling the resolutions -- 240p (300 Kbps), 360p (500 Kbps), 480p (1000 Kbps), 720p (2000 Kbps), 1080p (3000 Kbps), and 1440p (6000 Kbps). As each trace file is of a different length, we have also modified the streaming sessions of the different clients. For example, the Lumos-5G traces have a streaming session of 110 seconds, Irish traces of 120 seconds, MNWild-Ver of 130 seconds, and MNWild-TSA of 150 seconds. Each live video is encoded into 1-second segments. Each segment is further broken down into five chunks of 200 milliseconds. Other parameters, such as the client playback threshold and maximum latency threshold, are set to two seconds and five seconds, respectively. It is further assumed that when a user joins the session, it can request at most three video segments, and the streaming starts when it has downloaded at least two segments. The coefficients $\mu_1, \ \mu_2, \ \mu_3,\ \mu_4, \ \mu_5$ of (\ref{eq:qoelive}) are 0.2, 6.0, 1.0, 0.8, and 1.2, respectively.

\begin{figure}[!t]%
	\centering
        \includegraphics[width=0.75\columnwidth,keepaspectratio]{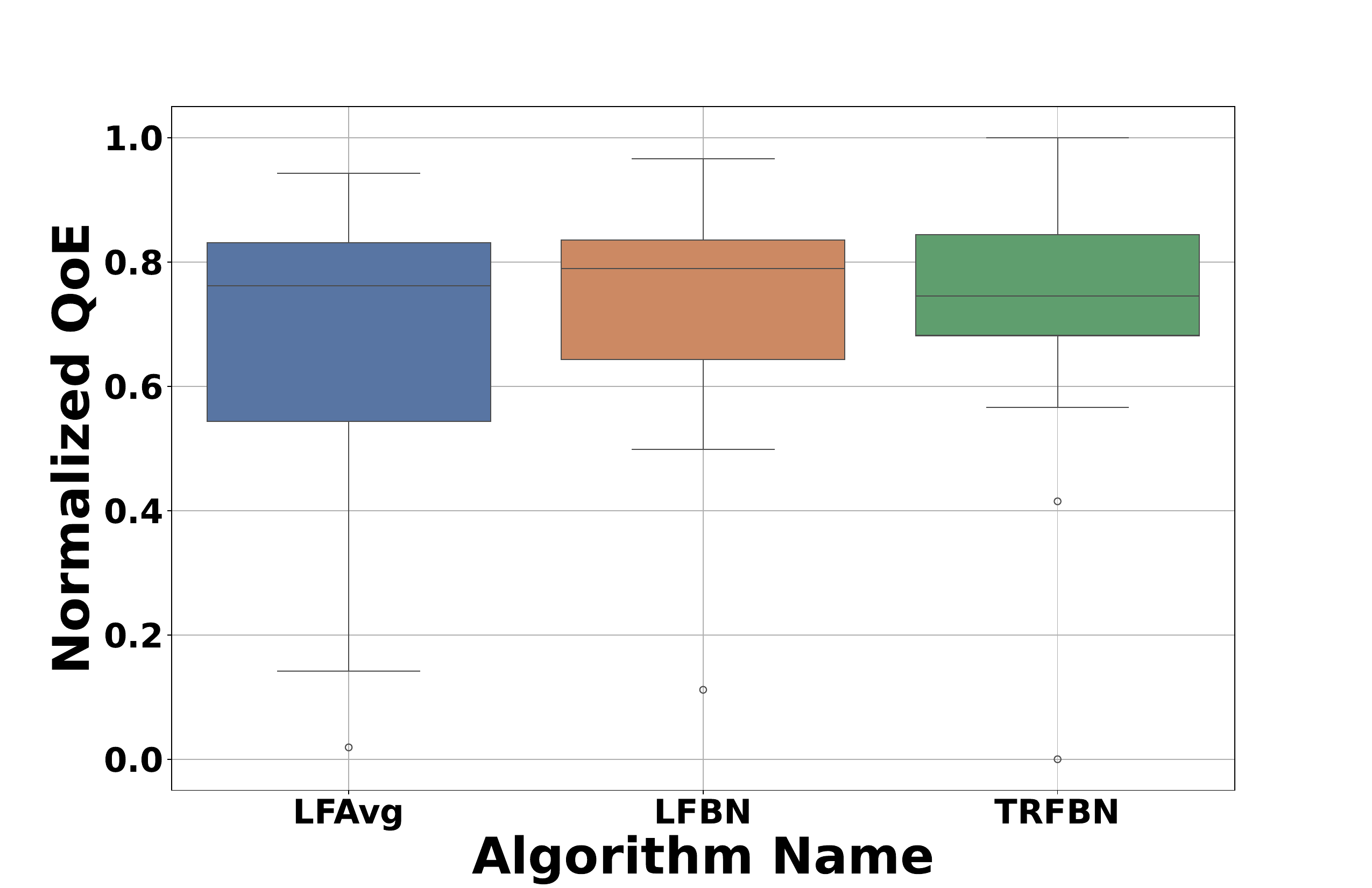}
	\caption{Live-Streaming QoE with the Federated throughput prediction algorithms LSTM+\fedbn{} (LFBN), Transformers+\fedbn{} (TRFBN) and the baseline LSTM+\fedavg{} (LFAvg). Number of Cohorts = 20, H = 15 secs. F = 1 sec.}
	\label{fig:BenchQoE}
\end{figure}
\begin{figure*}[h]
\centering
\begin{subfigure}{0.25\textwidth}
         \centering
         \includegraphics[width=\columnwidth,keepaspectratio]{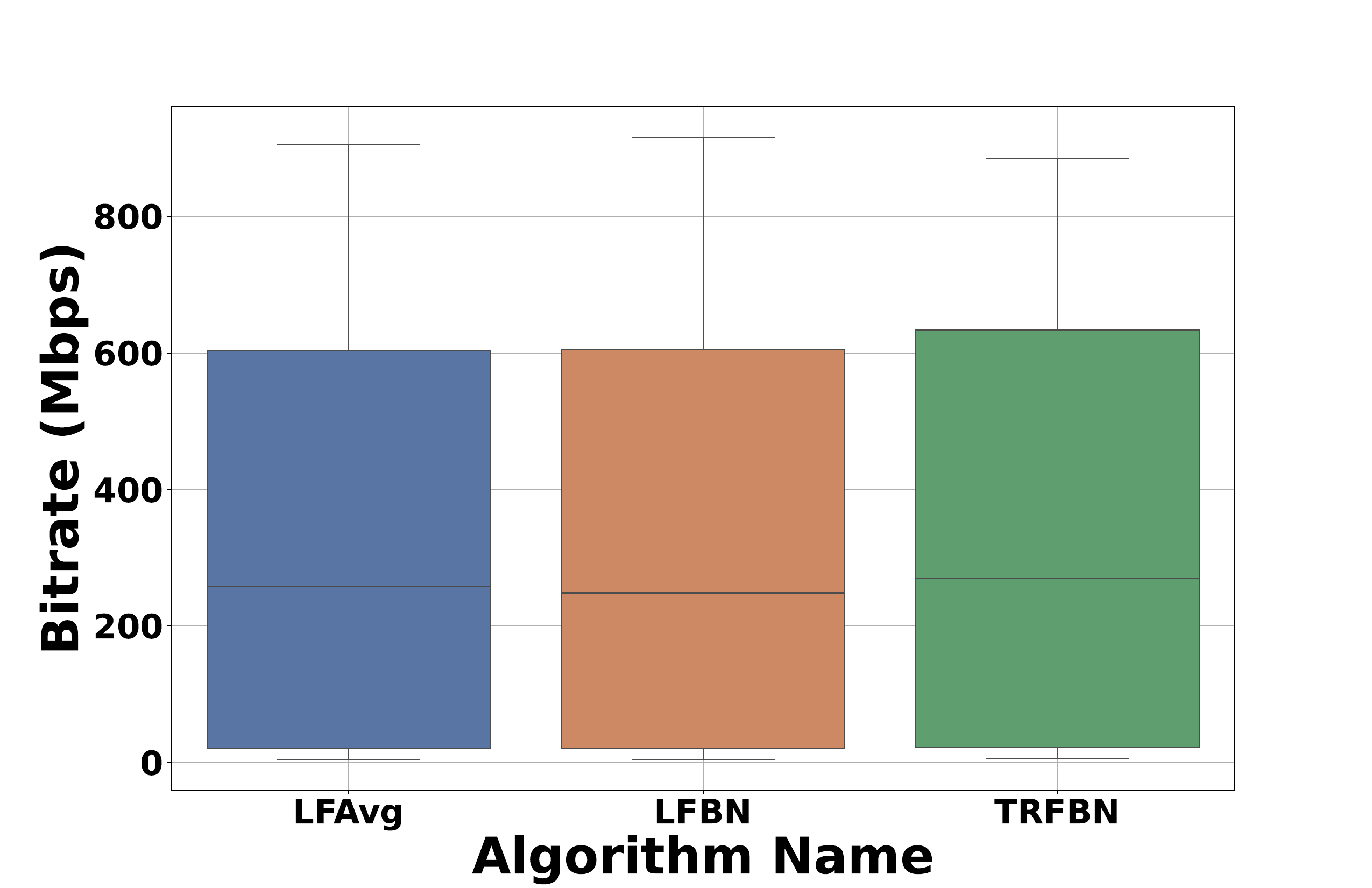}
\caption{Bitrates.}
\label{fig:Bench_Bitrate}
     \end{subfigure}
 \hspace*{-0.5cm}       
     \begin{subfigure}{0.25\textwidth}
         \centering
         \includegraphics[width=\columnwidth,keepaspectratio]{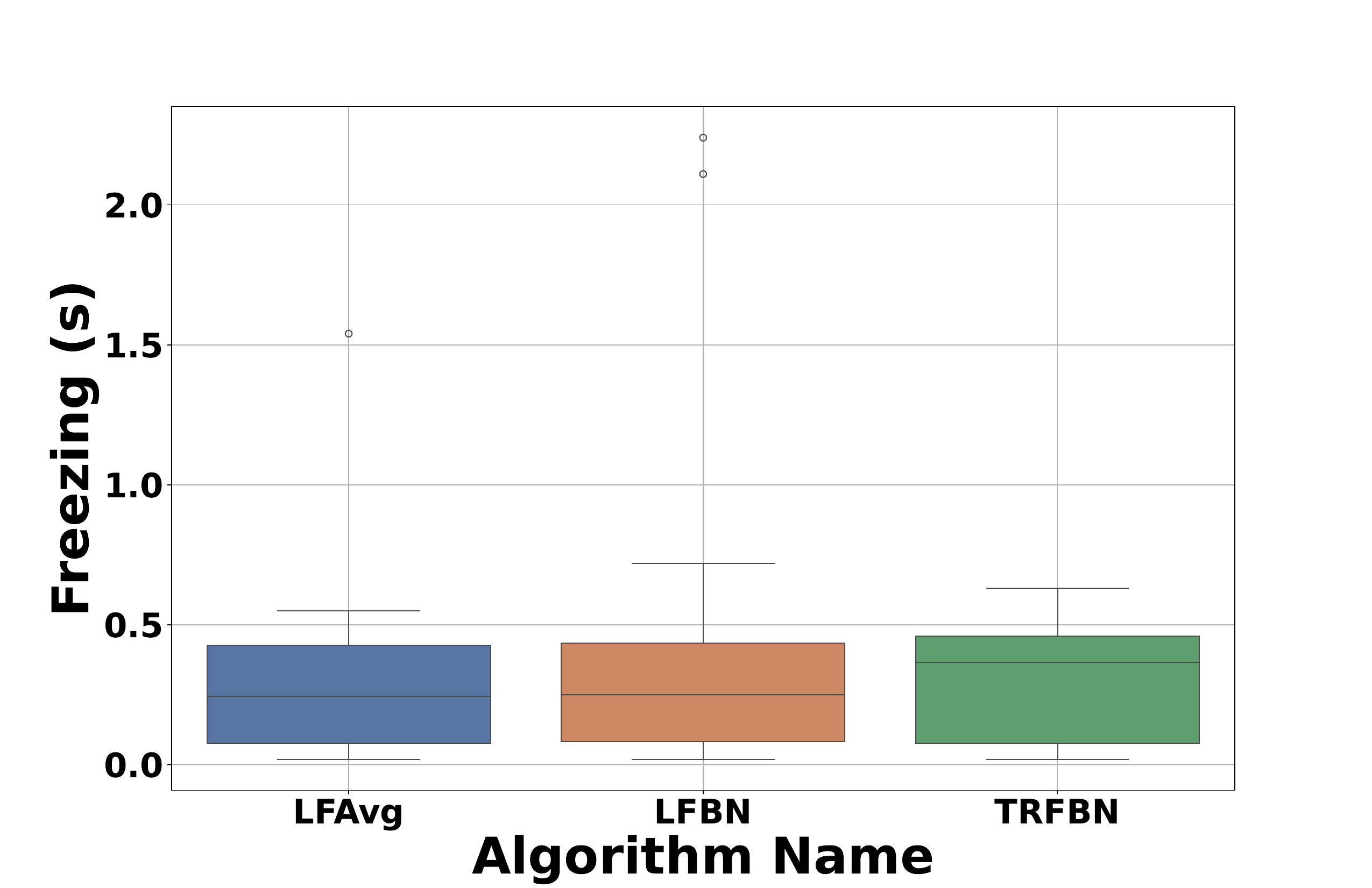}
\caption{Freezing Time.}
\label{fig:Bench_Freezing}
     \end{subfigure}
 \hspace*{-0.5cm}     
     \begin{subfigure}{0.25\textwidth}
         \centering
         \includegraphics[width=\columnwidth,keepaspectratio]{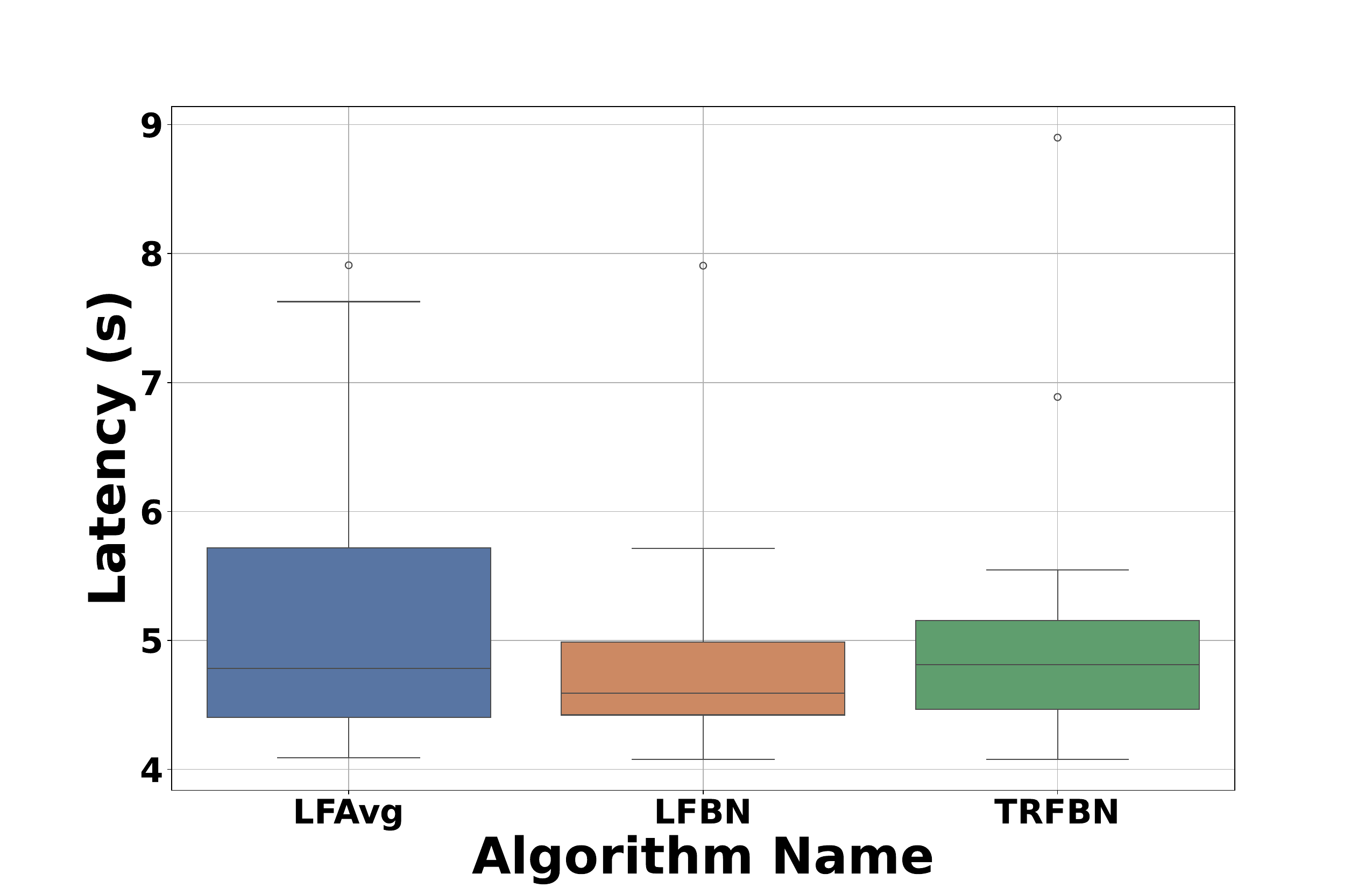}
	\caption{Latency.}
	\label{fig:Bench_latency}
     \end{subfigure}
      \hspace*{-0.5cm}       
     \begin{subfigure}{0.25\textwidth}
         \centering
\includegraphics[width=\columnwidth,keepaspectratio]{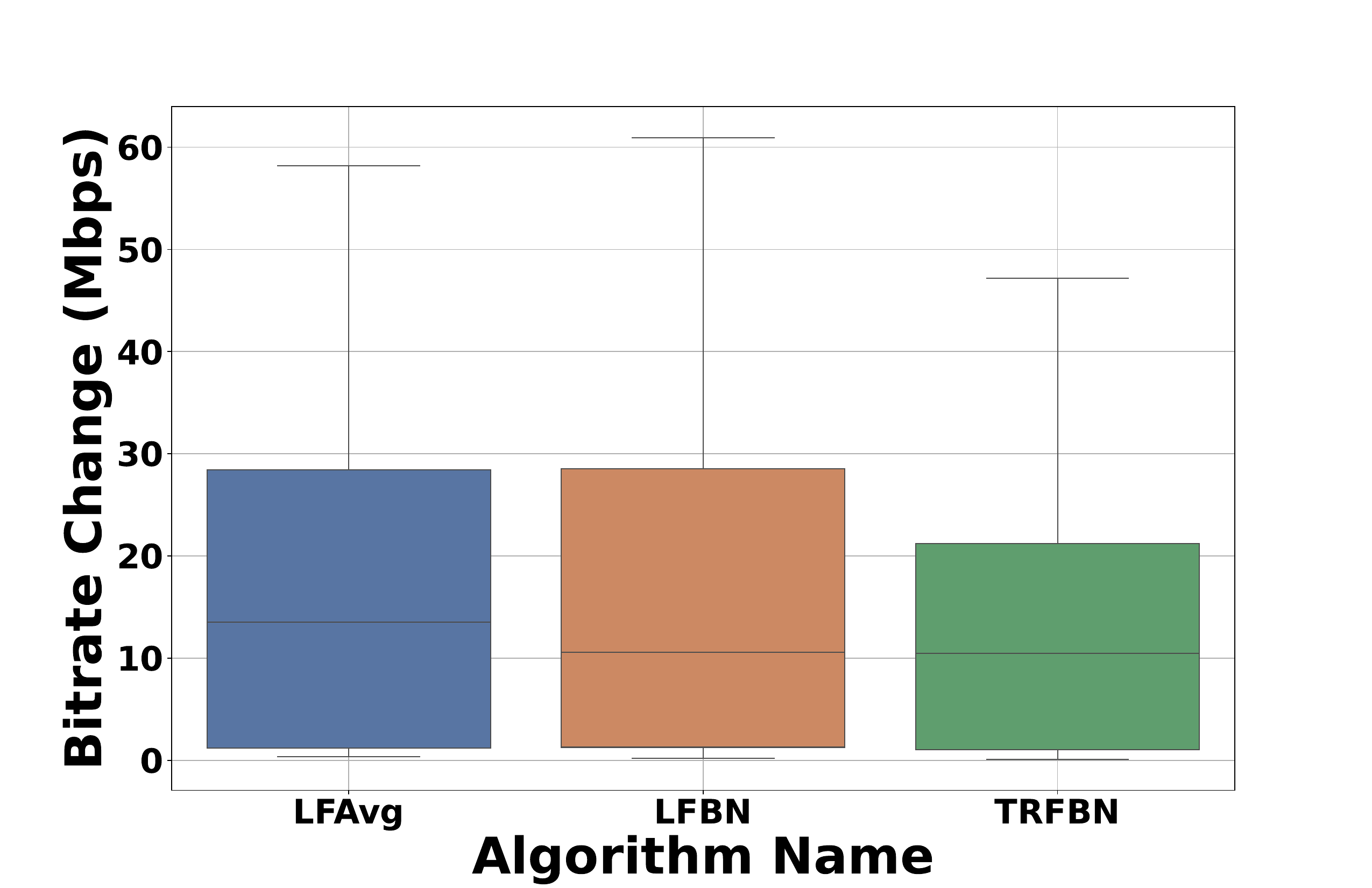}
	\caption{Bitrate Change.}
	\label{fig:Bench_bitrate_change}
     \end{subfigure}
        \caption{Performance of the individual Live Streaming QoE components. Number of Cohorts = 20, H = 15 secs, F = 1 sec.}
        \label{fig:QoEcomponents}
\end{figure*}

\subsubsection{\textbf{Results}}
We have used the outcome of the live-streaming protocol to compare the performance of the two federated throughput prediction algorithms - LSTM+\fedbn{} and Transformers+\fedbn{} (TRFBN), with the baseline LSTM+\fedavg{}~\cite{Sen2023}. As before, we have considered the history and prediction windows to be of $H=15$ seconds and $F=1$ second, respectively. 

\figurename~\ref{fig:BenchQoE} shows the \ac{QoE} of the three algorithms. It may be noted that the mean QoE delivered by LSTM+\fedavg{}, LSTM+\fedbn{}, and Transformer+\fedbn{} are 0.631, 0.705, and 0.702, respectively. This is in corroboration with our earlier inference that \fedbn{} is more adept at capturing the data heterogeneity than \fedavg{}. It is for this reason that LSTM+\fedavg{} also shows a higher variance than LSTM+\fedbn{} and Transformer+\fedbn{}.  

A deeper dive into the individual \ac{QoE} parameters, as illustrated in \figurename~\ref{fig:QoEcomponents}, reveals that although the three algorithms give comparable bitrates and freezing times, the latency offered by the \fedbn{} aggregation strategies, especially when combined with LSTM is the lowest of all. This can be attributed to the high accuracy yielded by LSTM as well as \fedbn{}'s ability to better handle heterogeneity. An important point to note is that the LSTM+\fedavg{} algorithm has a higher variance, often violating the five-second latency threshold. 
From the results, it may be inferred that FL-based throughput prediction performs considerably well to deliver a high QoE to live-streaming users. It may also be inferred that both LSTM and Transformers are robust to the heterogeneity in 5G, when the historical window size is higher. Moreover, of the different aggregation strategies, \fedbn{} has been found to handle the heterogeneity best. In the next section, we consolidate the key takeaways of our benchmarking study.  
\section{Key Takeaways and Future Directions}\label{sec:keytakeaway}
\subsection{Key Takeaways}
The key takeaways from the benchmarking study along with the live-streaming PoC are as follows. 
\label{sec:key_take}
\newadd{
\begin{enumerate}
    \item \textbf{Efficacy of LSTM and Transformer Architectures:} 
    Among the four evaluated architectures (CNN, \ac{LSTM}+\ac{CNN}, LSTM, Transformer), both LSTM and Transformer models consistently achieved high predictive accuracy. In contrast, CNN and hybrid \ac{LSTM}+\ac{CNN} models underperformed due to their limited capacity to capture long-range temporal dependencies in throughput data. This demonstrates the superiority of recurrent and attention-based models for time-series prediction in federated environments.
    \item \textbf{Trade-off Between Convergence Speed and Stability:} 
    Although LSTM and Transformer models exhibited comparable final $R^2$ scores, Transformers achieved faster convergence. However, Transformer models demonstrated higher variance in the $R^2$ score 
across clients, suggesting that while training rounds are more efficient, they may be less stable in heterogeneous environments.
    \item \textbf{Robustness of \fedbn{} Across Model Families:} 
    Among the aggregation strategies considered, \fedbn{} consistently delivered improved or comparable accuracy, particularly for CNN and \ac{LSTM}+\ac{CNN} models. For LSTM and Transformer architectures, all aggregation strategies yielded similar performance, though \fedbn{} exhibited marginally lower variance. These results highlight \fedbn{} as a strong candidate for deployment in non-IID federated settings.
    \item \textbf{Influence of Historical Window Size on Forecasting Accuracy:} 
    Parametric analysis revealed that Transformer models require longer historical input sequences to achieve high $R^2$ score, whereas LSTM models reach similar accuracy with shorter history lengths. This suggests that LSTM models are more efficient in leveraging short-term dependencies, making them more suitable for deployment on resource-constrained edge devices.
    \item \textbf{Impact of Cohort Size and Client Diversity:} 
    The performance of federated models varied with the size and composition of client cohorts. Larger cohort sizes led to increased variance, particularly when clients were drawn from diverse datasets with differing network conditions. This asserts the importance of careful cohort selection in federated deployments.
    \item \textbf{Impact on Practical QoE Metrics in Live Streaming Applications:} Our live-streaming PoC study reveals that federated throughput prediction, especially when combined with \fedbn{}, significantly improves user-perceived QoE. Notably, the combination of LSTM+\fedbn{} achieves the lowest average latency and least bitrate variation while maintaining competitive bitrate and freezing time metrics. This suggests that accurate and personalized throughput prediction via FL improves the end-to-end prediction accuracy and translates to tangible improvements in real-time adaptive video streaming.
\end{enumerate}}
\subsection{Future Directions}
\label{sec:future}
\newadd{This study provides a comprehensive benchmarking analysis of federated learning (FL) strategies for throughput prediction in 5G/6G networks. While the evaluation spans various model architectures and aggregation methods, it remains limited to simulations over publicly available datasets. A natural extension is the deployment of a real-world testbed involving physical mobile devices connected over heterogeneous cellular networks. Such a setup would allow for the empirical validation of FL strategies under more realistic network dynamics, hardware variability, and device-level constraints such as energy consumption and thermal throttling.}

\newadd{Another important avenue for future exploration is the consideration of \emph{model heterogeneity} across clients. The current study assumes a uniform model architecture across all participating devices, which may not be feasible given the computational and memory disparities among mobile hardware~\cite{mo2025enhancingefficiencymultidevicefederated}. Investigating federated distillation~\cite{mora2024knowledge} and split learning~\cite{mo2025enhancingefficiencymultidevicefederated} could help enable effective FL deployment in resource-diverse environments. Finally, extending the benchmarking framework to account for communication efficiency~\cite{10494348} can further align FL throughput prediction systems with the constraints of practical deployments in next-generation networks.}

\section{Conclusion}\label{sec:conclusion}

This work presents a novel benchmarking framework for FL-based throughput prediction tailored to heterogeneous 5G networks, with a specific focus on live-streaming applications. Unlike prior efforts, we systematically compare multiple FL aggregation strategies (\textit{FedAvg}, \textit{FedProx}, and \textit{FedBN}) across deep learning architectures (CNN,  LSTM+CNN, LSTM, and Transformer), providing the first comprehensive evaluation of how model-strategy combinations perform under varying levels of system and statistical heterogeneity. Our results show that LSTM and Transformer models consistently outperform CNN-based baselines while maintaining aggregation-strategy-agnostic performance, with mean R² scores exceeding 0.93 across all cohorts. We find that while Transformers converge in half as many rounds as LSTMs, they require longer historical windows to match LSTM performance. This suggests a trade-off- positioning Transformers for real-time inference, and LSTM for scenarios demanding stability and lower overhead.

A key contribution is our integration of FL-driven throughput forecasting into a live-streaming proof-of-concept using MPC-based bitrate adaptation. We observe an 11.7\% QoE improvement with the LSTM+\textit{FedBN} model over LSTM+\textit{FedAvg}. Our current implementation relies on emulation over publicly available 5G datasets and assumes homogeneous model architectures across clients. Future work will explore deployment on real mobile testbeds, client-specific model heterogeneity via model distillation or split learning, and improved communication efficiency for edge FL.

\bibliographystyle{IEEEtran}
\bibliography{Fedput2_Main}

\begin{IEEEbiography}
[{\includegraphics[width=1in,height=1.25in,clip,keepaspectratio]{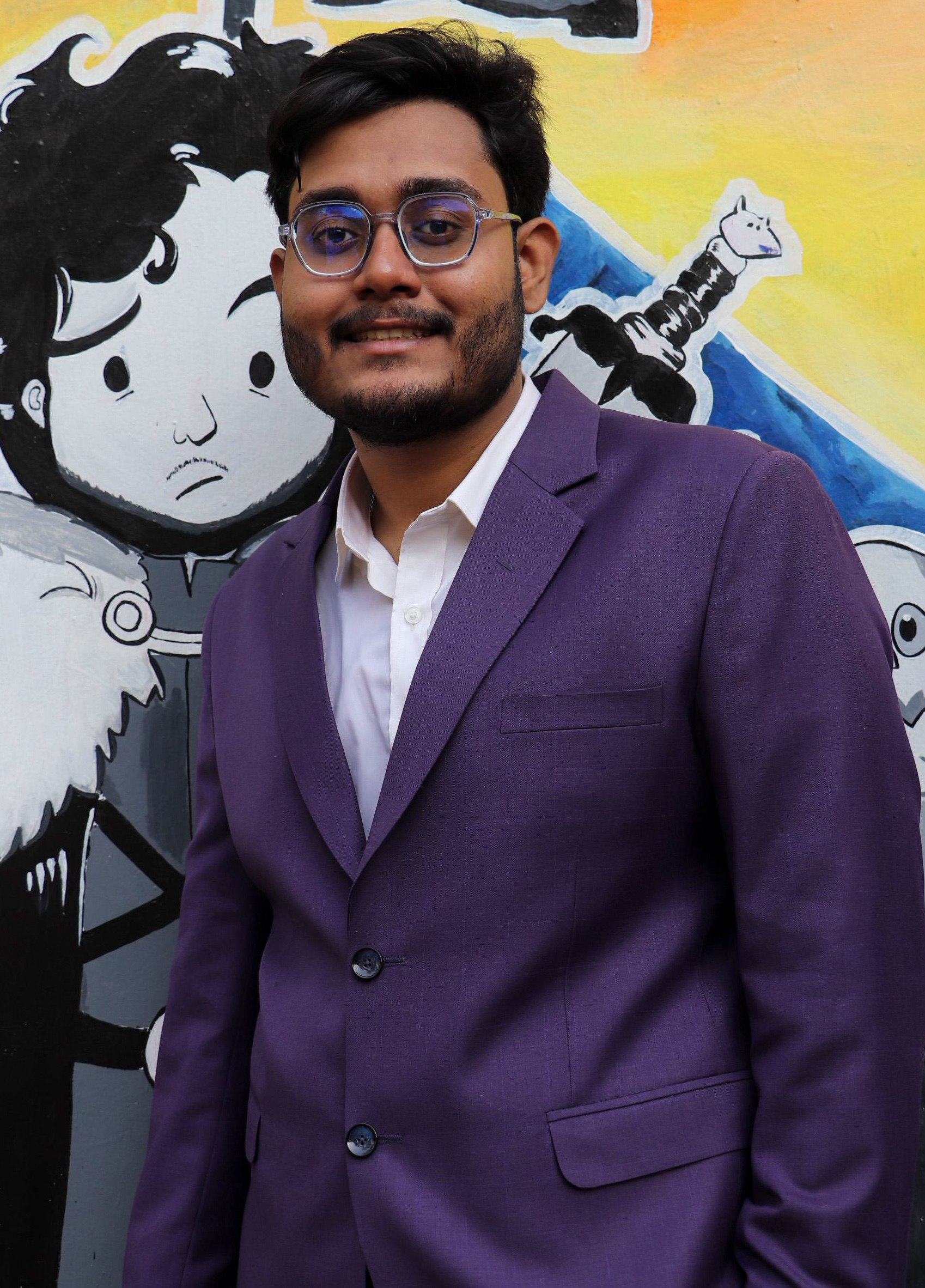}}]{Yuvraj Dutta}
 has obtained his B.Tech degree from the National Institute of Technology Rourkela, India, in Electronics and Communication Engineering, following  his Higher Secondary level education from Notre Dame College, Bangladesh. Originally from Bangladesh, Yuvraj is the recipient of the prestigious Indian Council for Cultural Relations (ICCR) scholarship provided by the Govt. of India, which enabled him to receive his BTech degree from India. His research interests lie in the intersection of AI-based communication systems, signal processing, and image processing.
\end{IEEEbiography}
\begin{IEEEbiography}
[{\includegraphics[width=1in,height=1.25in,clip,keepaspectratio]{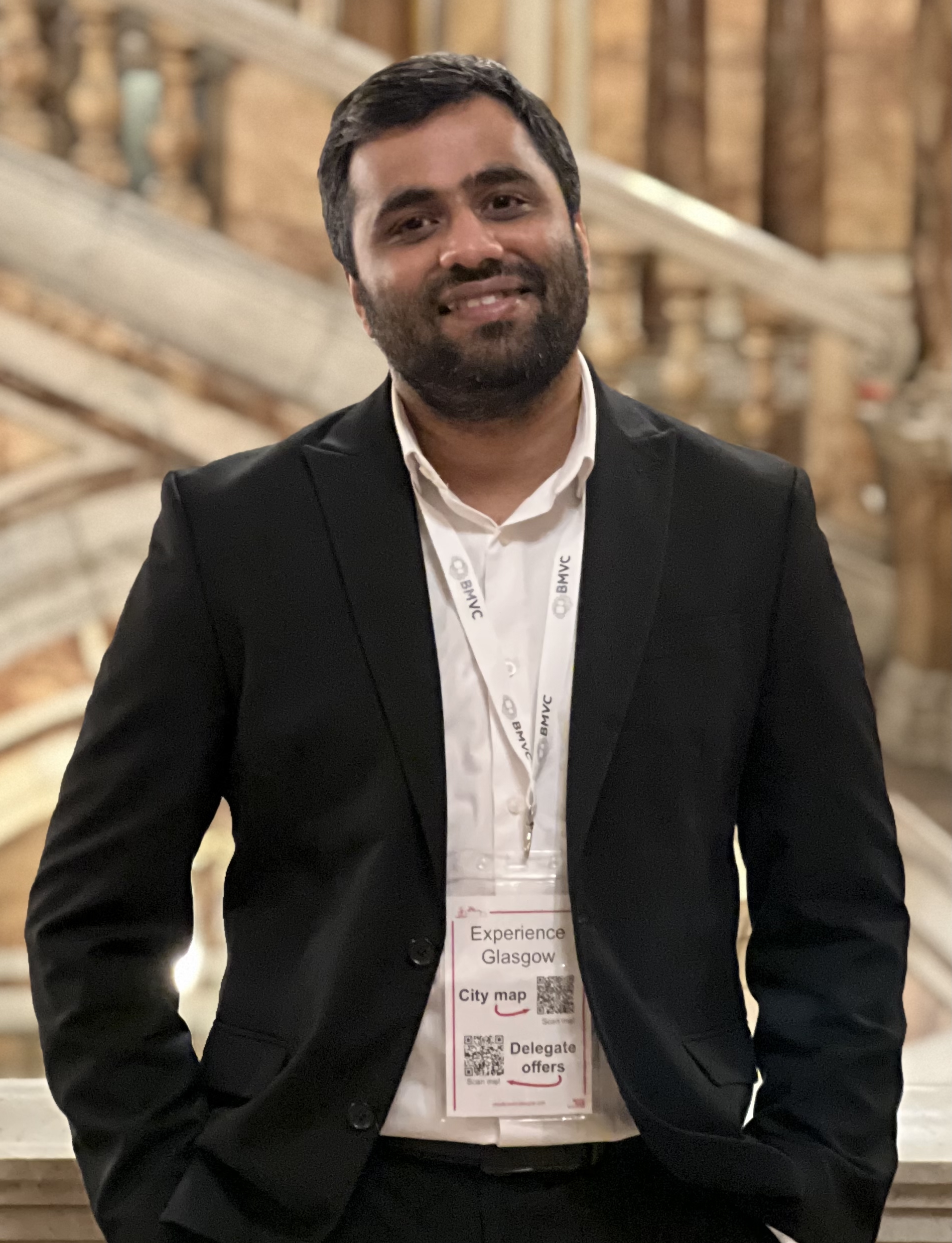}}]{Soumyajit Chatterjee} (Member, IEEE) is a Senior Research Scientist at Nokia Bell Labs, UK, and a Visiting Scholar at the University of Cambridge. He received his Ph.D. in Computer Science from IIT Kharagpur in 2022, where he worked on developing automated annotation frameworks with multimodal sensing approaches. His research interests include federated learning, on-device adaptation, label-efficient learning, and human-centered AI for pervasive computing. He has authored several publications in premier conferences and journals and has filed multiple patents in efficient ML systems. Dr. Chatterjee regularly mentors graduate interns and has actively contributed to organizing academic workshops and reviewing for IEEE and ACM venues.
\end{IEEEbiography}
\begin{IEEEbiography}[{\includegraphics[width=1in,height=1.25in,clip,keepaspectratio]{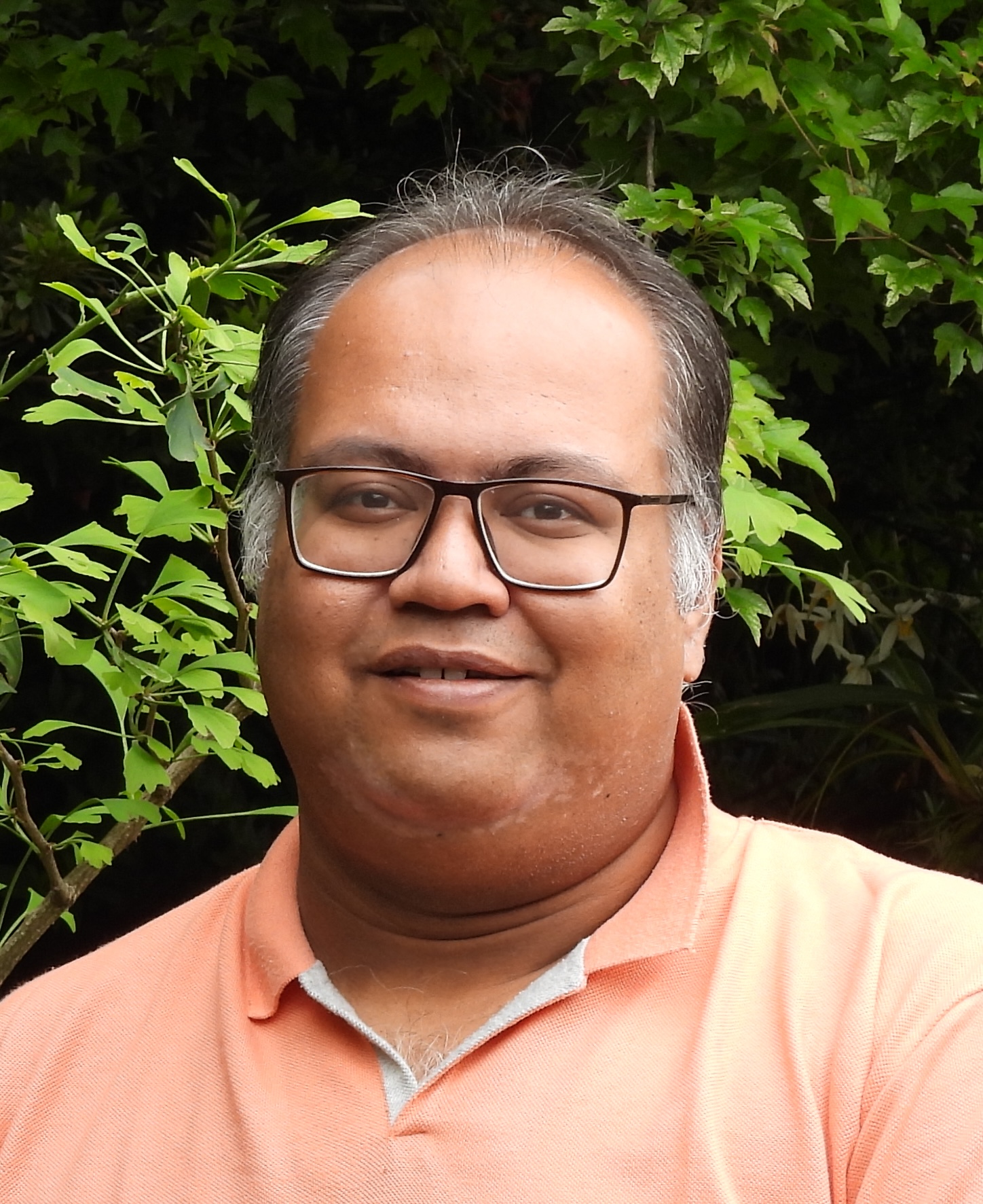}}]{ Sandip Chakraborty} (Senior Member, IEEE) is working as an Associate Professor in the Department of Computer Science and Engineering at the Indian Institute of Technology Kharagpur, and leading the Ubiquitous Networked Systems Lab (UbiNet, \url{https://ubinet-iitkgp.github.io/ubinet/}). He obtained his Bachelor’s degree from Jadavpur University, Kolkata in 2009 and M Tech and Ph.D., both from IIT Guwahati, in 2011 and 2014, respectively. The primary research interests of Dr. Chakraborty are Distributed Systems, Pervasive, Ubiquitous and Edge Computing, and Human-Computer Interactions. He is working as an Area Editor of IEEE Transactions on Services Computing, Elsevier Ad Hoc Networks and Elsevier Pervasive and Mobile Computing journals. 
\end{IEEEbiography}
\begin{IEEEbiography}
[{\includegraphics[width=1in,height=1.25in,clip,keepaspectratio]{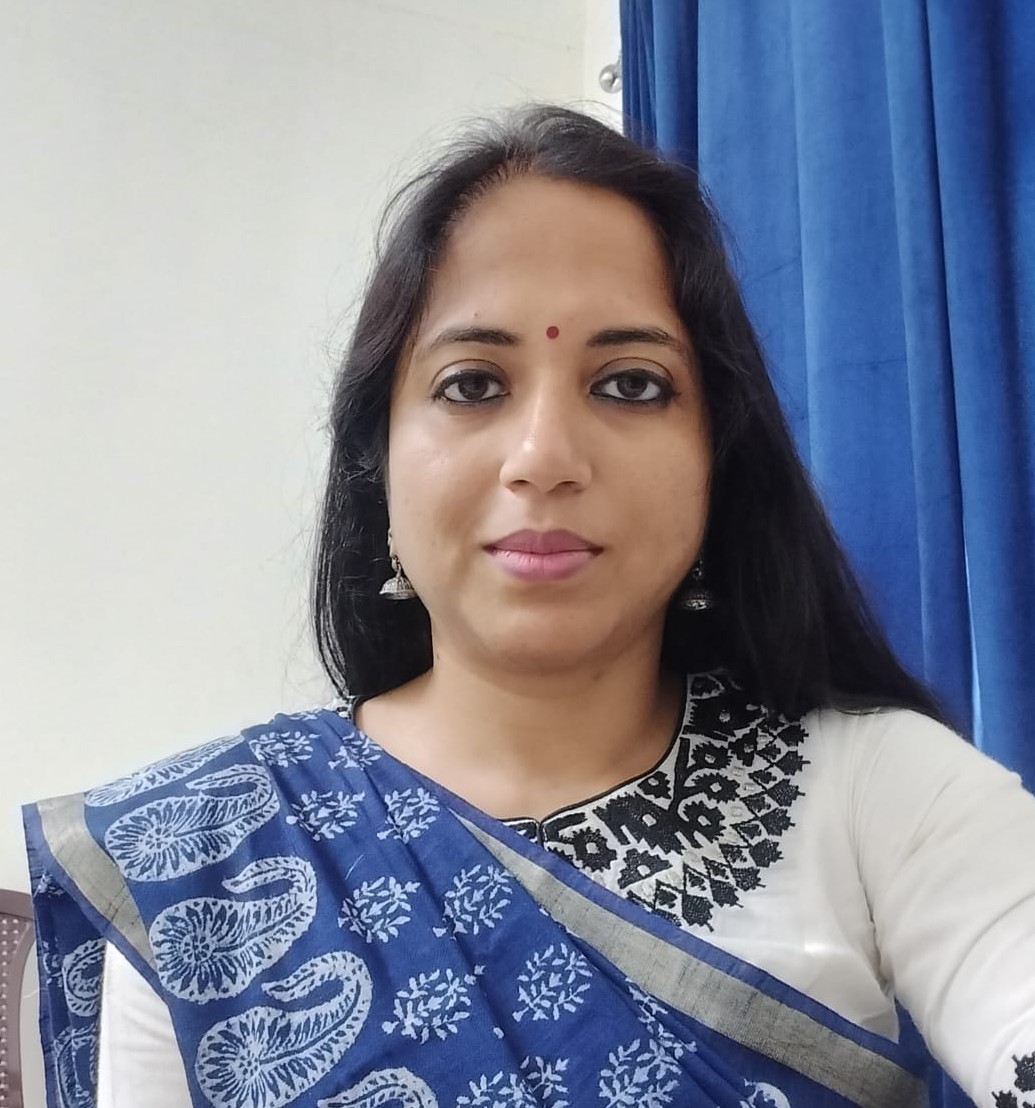}}]{Basabdatta Palit}
 has received her Ph.D. Degree from IIT Kharagpur, India in 2018. She completed her B.Tech from West Bengal University of Technology in 2009 and her M.Tech. degree in 2011 from Jadavpur University, Kolkata, India. She is currently an Assistant Professor in the Department of Electronics and Communication Engineering at National Institute of Technology, Rourkela, India. Her research interests include Cross-Layer Design and Modeling, Application of ML in Communication, Joint Sensing and Communication, URLLC, etc.
\end{IEEEbiography}

\begin{acronym}
	\acro{2G}{2$^\mathrm{nd}$ Generation}
	\acro{3G}{3$^\mathrm{rd}$ Generation}
	\acro{4G}{4$^\mathrm{th}$ Generation}
	\acro{5G}{5$^\mathrm{th}$ Generation}
	\acro{A3C}{Actor-Critic}
	\acro{ABR}{adaptive bitrate}
	\acro{AR}{Augmented Reality}
	\acro{ARE}{Absolute Value of Residual Error}
        \acro{ARMA}{Auto Regressive Moving Average}
	\acro{ARIMA}{Auto Regressive Integrated Moving Average}
	\acro{BLUE}{Best Linear Unbiased Estimator}
	\acro{BS}{Base Station}
	\acro{CDN}{Content Distribution Network}
 \acro{CNN}{Convolutional Neural Networks}
	\acro{CQI}{Channel Quality Index}
	\acro{D1} {Data set from \cite{Raca2020_1}}
	\acro{D2} {Data set from \cite{Narayanan2021}}
	\acro{DASH}{Dynamic Adaptive Streaming over HTTP}
	\acro{DTR}{Decision Tree Regression}
	\acro{DRX}{Discontinuous Reception}
	\acro{DL}{Deep Learning}
	\acro{DT}{Decision Trees}
        \acro{MPC}{Model Predictive Control}
    \acro{EB}{Exabyte}
    \acro{FFNN}{Feed Forward Neural Network}
        \acro{FMPC}{Fast \ac{MPC}}
        \acro{FL}{Federated Learning}
        \acro{FedAvg}{Federated Averaging}
	\acro{KF}{Kalman Filter}
        \acro{KFTP}{Kalman Filter based Throughput Prediction}
        \acro{KNNR}{K-Nearest Neighbours Regression}
        \acro{MA}{Moving Average}
	\acro{MAE}{Mean Absolute Error}
	\acro{mn}{Measurement Noise}
	\acro{MSE}{Mean Squared Error}
	\acro{ML}{Machine Learning}
	\acro{MLP}{Multilayer Perceptron}
	\acro{MLR}{Multiple Linear Regression}
 \acro{MR}{Mixed Reality}
	\acro{EDGE}{Enhanced Data Rates for \ac{GSM} Evolution.}
	\acro{eNB}{evolved NodeB}
    \acro{GBDT}{Gradient Boosting Decision Trees}
	\acro{GSM}{Global System for Mobile}
	\acro{HD}{High Definition}
	\acro{HSPA}{High Speed Packet Access}
        \acro{IoT}{Internet of Things}
        \acro{KNN}{K-Nearest Neighbours}
	\acro{LTE}{Long Term Evolution}	
	\acro{LSTM}{Long Short Term Memory}
    \acro{LCNN}{LSTM+CNN}
	\acro{NR}{New Radio}
	\acro{pn}{Prediction Noise}
	\acro{HVPM}{High voltage Power Monitor}
	\acro{gNB}{gNodeB}
	\acro{GPX5} {Google Pixel 5}
    \acro{KDE}{Kernel Density Estimate}
    \acro{NAA}{Network Aware Applications}
    \acro{PDF}{probability density function}
	\acro{QoS}{Quality of Service}
	\acro{QoE}{Quality of Experience}
	\acro{RF}{Random Forest}
	\acro{RFL}{Random Forest Learning}
	\acro{RL}{Reinforcement Learning}
	\acro{RMSRE}{Root Mean Square Relative Error}
	\acro{RNN}{Recurrent Neural Network}
	\acro{RRC}{Radio Resource Control}
    \acro{RSS}{Residual Sum of Squares}
	\acro{RSSI}{Received Signal Strength Indicator}
	\acro{RSRP}{Reference Signal Received Power}
	\acro{R2}{R-squared metric}
	\acro{RSRQ}{Reference Signal Received Quality}
        \acro{SGS20U}{Samsung Galaxy S20 Ultra}
	\acro{SA}{Standalone 5G}
	\acro{SER}{Signal to Error Ratio}
	\acro{SVR}{Support Vector Regressor}
	
	\acro{NSA} {Non-standalone 5G}
 \acro{RMPC}{Robust MPC}
	\acro{SGS20U} {Samsung Galaxy S20 Ultra 5G}
    \acro{SGS10} {Samsung Galaxy S10 5G}

	\acro{SINR}{signal-to-interference-plus-noise-ratio}
	\acro{SNR}{signal-to-noise-ratio}
 \acro{TCP}{Transmission Control Protocol}
	\acro{TNSA}{TMobile, NSA+LTE}
	\acro{TSA}{TMobile, SA}
        \acro{THPT}{Throughput}
        \acro{EWMA}{Exponential Weighted Moving Average}
	\acro{UE}{User Equipment}
	\acro{UHD}{Ultra HD}
    \acro{UAV}{Unmanned Aerial Vehicles}
	\acro{Ver}{Verizon, Default}
	\acro{VoLTE}{Voice over LTE}
	\acro{VR}{Virtual Reality}
        \acro{WCDMA}{Wideband Code Division Multiple Access}
	\acro{WiFi}{Wireless Fidelity}
	\acro{XGBoost}{Extreme Gradient Boost}
	\acro{SUMO}{Simulation of Urban MObility}
	\acro{MCS}{Modulation Coding Scheme}
        \acro{HM}{Harmonic Mean}
        \acro{VoD}{video-on-demand}
\end{acronym}
\end{document}